\documentclass[aps,pra,11pt]{revtex4-2}
\usepackage{amsthm}  
\usepackage{amsmath,amssymb,amsfonts,graphicx,bbm,mathtools,braket,color,dsfont}
\usepackage{array,dcolumn,mathrsfs}
\usepackage{caption,subcaption}
\usepackage{diagbox,natbib,xcolor,graphicx} 

\newtheorem{theorem}{Theorem}[section]
\newtheorem{lemma}[theorem]{Lemma}
\newtheorem{definition}{Definition}[section]
\newcommand{\half}{\frac{1}{2}}
\newcommand{\op}[2]{\ket{#1}\!\!\bra{#2}}

\newcommand{\ip}[2]{\left\langle{#1}\right|\left.\!{#2}\right\rangle}
\DeclareMathOperator{\Tr}{Tr}

\newcommand{\SLD}{\mathrm{SLD}}

\begin{document}
\title{Weight-dependent and weight-independent measures of quantum incompatibility in multiparameter estimation}
\author{Jiayu He} 
\email{jiayu.he@helsinki.fi (J. He)}
\affiliation{QTF Centre of Excellence, Department of Physics, University of Helsinki, FI-00014 Helsinki, Finland}
\author{Gabriele Fazio}
\email{gabriele.fazio@studenti.unimi.it (G. Fazio)}
\affiliation{Dipartimento di Fisica Aldo Pontremoli, Università degli Studi di Milano, I-20133 Milano, Italy}
\author{Matteo G. A. Paris}
\email{matteo.paris@fisica.unimi.it (M. G. A. Paris)}
\affiliation{Dipartimento di Fisica Aldo Pontremoli, Università degli Studi di Milano, I-20133 Milano, Italy}

\begin{abstract}
Multiparameter quantum estimation faces a fundamental challenge due to the inherent incompatibility of optimal measurements for different parameters, a direct consequence of quantum non-commutativity. This incompatibility is quantified by the gap between the symmetric logarithmic derivative (SLD) quantum Cram\'er-Rao bound, which is not always attainable, and the asymptotically achievable Holevo bound. This work provides a comprehensive analysis of this gap by introducing and contrasting two scalar measures. The first is the weight-independent quantumness measure $R$, which captures the intrinsic incompatibility of the estimation model. The second is a tighter, weight-dependent measure $T[W]$ which explicitly incorporates the cost matrix $W$ assigning relative importance to different parameters. We establish a hierarchy of bounds based on these two measures and derive necessary and sufficient conditions for their saturation. Through analytical and numerical studies of tunable qubit and qutrit models with SU(2) unitary encoding, we demonstrate that the weight-dependent bound $C_{T}[W]$ often provides a significantly tighter approximation to the Holevo bound $C_{H}[W]$ than the $R$-dependent bound, especially in higher-dimensional systems. We also develop an approach based on $C_{T}[W]$ to compute the Holevo bound $C_{H}[W]$ analytically. Our results highlight the critical role of the weight matrix's structure in determining the precision limits of multiparameter quantum metrology.
\end{abstract}
\maketitle

\section{Introduction}

The precise estimation of unknown parameters is a cornerstone of science and technology. Quantum metrology exploits quantum-mechanical effects, such as entanglement and squeezing, to achieve a precision surpassing the limits of classical strategies\cite{giovannetti2004,GAIBA2009934,kruse2016,demkowicz2014,riedel2010,PhysRevLett.116.053601,pezze2018,giovannetti2011,schnabel2010,bina2016dicke,montenegro2025quantum,montenegro2025heisenberg}. While single-parameter estimation is well-understood, many practical applications require the simultaneous estimation of multiple parameters\cite{defienne2024,magana2019,altmann2018,degen2017,abbott2016,adhikari2014,proctor2018,zhuang2018,zaiser2016,guo2020,lawrie2019,goda2008,tse2019}.  In turn, multiparameter estimation can offer significant advantages, including resource efficiency and the avoidance of systematic errors associated with separate calibration\cite{Humphreys2013,yue2014,liu2016,gagatsos2016,szczykulska2016,gf25}.

However, multiparameter quantum estimation introduces a unique challenge not present in the classical or single-parameter quantum case: the problem of \emph{incompatibility}\cite{zhu2015,heinosaari2016,ragy2016,Candeloro2024}. In quantum mechanics, the optimal measurement for one parameter may fail to commute with the optimal measurement for another. This non-commutativity implies that, in general, no single measurement strategy can extract maximal information about all parameters simultaneously. The fundamental limits of precision and the influence of non-commutativity are in turn captured by a family of bounds. The simplest is the \emph{quantum Cram\'er-Rao bound} (QCRB) based on the symmetric logarithmic derivative (SLD), denoted $C_{\mathrm{\SLD}}$\cite{braunstein1994}. While analytically tractable, this bound is not always attainable. The ultimate, asymptotically achievable limit is given by the natively scalar Holevo bound, $C_{{H}}$\cite{holevo2011}.

The gap between $C_{\SLD}$ and $C_{{H}}$ is a direct measure of the quantum incompatibility of the estimation problem. A key quantity capturing this incompatibility is the \emph{mean Uhlmann curvature} (MUC) matrix $U$, whose elements reflect the average non-commutativity of the parameter generators. A scalar measure of {\em quantumness} \cite{carollo2019}, defined as
$R = \left\lVert i Q^{-1} U \right\rVert_\infty$,
where $Q$ is the quantum Fisher information matrix, provides an intrinsic measure of incompatibility, independent of the relative importance assigned to different estimation errors.

In realistic scenarios, however, not all estimation errors are equally costly. This relative importance is encoded in a positive definite weight matrix $W$. A central question arises: how does the structure of $W$ influence the severity of the incompatibility problem? To address this, we introduce a weight-dependent measure, $T[W]$, which tightens the hierarchy of bounds by explicitly accounting for $W$. This leads to an intermediate bound
\[
C_{T}[W] = (1+T[W]) C_{\SLD}[W]\, ,
\]
which lies between Holevo bound $C_{H}[W]$ and the $R$-dependent bound
\[
C_{R}[W] = (1+R) C_{\SLD}[W]\, .
\]
In this paper, we present a systematic study of these weight-dependent and weight-independent measures of quantum incompatibility. We derive the fundamental relations between the bounds $C_{\SLD}[W], C_{H}[W], C_{{T}}[W]$, and $C_{{R}}[W]$, and establish conditions under which they coincide. We then explore these concepts in concrete physical settings. We analyze a versatile two-parameter qubit model where incompatibility can be tuned via an intermediate rotation, and we extend our analysis to a three-parameter qutrit model with $\mathrm{SU}(2)$ unitary encoding. Our results show that $C_{T}[W]$ typically provides a much tighter approximation to the challenging-to-compute Holevo bound $C_{{H}}[W]$ than $C_{{R}}[W]$, especially as the dimension of the system grows, emphasizing the practical relevance of the weight matrix in assessing metrological performance.

The paper is organized as follows: Section~II reviews the framework of multiparameter quantum estimation. Section~III introduces the quantumness measure $R$ and the weight-dependent measure $T[W]$, deriving their properties and the associated bounds. Section~IV discusses the effect of different weight matrix structures. Section~V analyzes a tunable two-parameter qubit model, and Section~VI investigates multiparameter estimation for $\mathrm{SU}(2)$ unitary encoding in both qubit and qutrit systems. Finally, we close the paper with a discussion of our findings and some concluding remarks. 

\section{Framework of Multiparameter quantum estimation}
In this Section, we outline the theoretical background for multiparameter quantum estimation. Let $\rho_{\vec{\lambda}}$ be a quantum state on a finite-dimensional Hilbert space, parameterized by a vector of $d$ real parameters $\vec{\lambda}=(\lambda_1,\ldots, \lambda_d)^T$, and $\left\{\Pi_k\right\}$ a positive operator-valued measurement (POVM)  with $\Pi_k\geq0$ and $\sum_{k} \Pi_k=\mathbb{I}$. The probability of obtaining outcome $k$ is given by $p_{\vec{\lambda}}(k) = \Tr \left[\rho_{\vec{\lambda}}\Pi_k\right]$. An 
estimator $\hat{\vec{\lambda}}(k)$ is the assigned to each outcome, and its performance is evaluated via the covariance matrix $\vec{V}(\hat{\vec{\lambda}})$, whose components read:
\begin{align*}
V_{\mu\nu} =&\sum_k p_{\vec{\lambda}}(k)[\lambda_\mu(k)-E_k(\hat \lambda_\mu)][ \lambda_\nu(k)-E_k(\hat \lambda_\nu)].
\end{align*}
where $E_k(\hat \lambda_\mu)$ denotes the expectation value of the estimator $\hat{\lambda}_\mu$ under the distribution $p_{\vec{\lambda}}(k)$.

Assuming locally unbiased estimators, i.e. $E[\hat{\lambda}_\mu] = \lambda_\mu$ and $\partial_\nu E(\hat{\lambda}_\mu) = \delta{\mu_\nu}$, the classical Cram\'er-Rao bound (CRB) provides a fundamental lower limit on the achievable covariance\cite{cramer1999}:
\begin{equation*}
 V(\hat{\vec{\lambda}})\geq \frac{1}{M}F^{-1}\,,
\end{equation*}
with $F$ the Fisher information matrix (FIM), and $M$ the number of measurement repetitions. The elements of FIM are defined as:
\begin{equation*}
F_{\mu\nu}= \sum_k \frac{\partial_\mu  p_{\vec{\lambda}}(k)\, \partial_\nu  p_{\vec{\lambda}}(k)}{p_{\vec{\lambda}}(k)}
\,.
\end{equation*}
The CRB can be saturated in the asymptotic limit of an infinite number of repeated experiments using Bayesian or maximum likelihood estimators \cite{kay1993}. 

In the quantum setting, due to non-commutativity of observables, multiple versions of quantum Fisher information matrices (QFIMs) arise. The most prominent among them are the symmetric logarithmic derivative (SLD) and the right logarithmic derivative (RLD) QFIMs, based on the operators $L_\mu^S$ \cite{helstrom1967}  and $L_\mu^R$ \cite{Yuen1973}respectively, which satisfy
\begin{align*}
\partial_\mu\rho_{\vec{\lambda}}=\frac{L_\mu^S\rho_{\vec{\lambda}}+\rho_{\vec{\lambda}} L_\mu^S}{2}\,, \;\;\partial_\mu \rho_{\vec{\lambda}}=\rho_{\vec{\lambda}} L_\mu ^R.
\end{align*}
Their associated QFIMs are
\begin{align*}
Q_{\mu \nu}:=\frac{1}{2}\Tr\left[\rho_{\vec{\lambda}}\{L_\mu^S,L_\nu^S\}\right]\,, \;\;
J_{\mu \nu}:=\Tr\left[\rho_{\vec{\lambda}} L_\mu^R L_\nu^{R\dagger}\right]\,.
\end{align*}
In the case of pure states $\rho_{\vec{\lambda}} = |\psi_{\vec{\lambda}}\rangle \langle \psi_{\vec{\lambda}}|$, the SLD QFIM reduces to
\begin{align*}
Q_{\mu\nu}  = 4\,\textrm{Re}\big(\ip{\partial_\mu\psi_{\vec{\lambda}}}{\partial_\nu\psi_{\vec{\lambda}}}-\ip{\partial_\mu\psi_{\vec{\lambda}}}{\psi_{\vec{\lambda}}}\ip{\psi_{\vec{\lambda}}}{\partial_\nu\psi_{\vec{\lambda}}}\big),
\end{align*}
where $\partial_k\equiv\partial_{\lambda_k}$.
\subsection{Quantum Cram\'er-Rao Bounds}  
Utilizing $Q $ and $J$, one can formulate scalar quantum Cram\'er-Rao bounds for estimation error under a given cost matrix $W$ (a real, positive definite $d \times d$ matrix), leading to
\begin{align*}
C_{\SLD}[W] & =\frac1M\,\Tr\left[ W\, Q^{-1}\right],\\
C_{\mathrm{RLD}}[W]& = \frac1M\,\left( \Tr\left[ W\, \textrm{Re}(J^{-1})\right]+\Tr \left[\left|\left|{ W\, \textrm{Im}(J^{-1}) }\right|\right|_1\right]\right)\,,
\end{align*}
where $||A||_1=\sqrt{A^\dag A}$ and $\textrm{Re}(A)$ and $\textrm{Im}(A)$ denote 
the real and imaginary parts of the complex-valued matrix $A$, respectively.
However, due to non-commuting SLDs, these bounds are not always tight. A more fundamental limit is provided by the Holevo bound\cite{holevo2011}:
\begin{equation*}
C_H[W]
=\min_{X \in \mathcal X} \Big\{\Tr\left[ W\, \textrm{Re}\left( Z[X]\right)\right]+\Tr \left[\left|\left|W\, \textrm{Im}\left( Z[ X]\right)\right|\right|_1\right]\Big\},
\end{equation*}
where $Z_{\mu\nu} := \Tr[\rho_{\vec{\lambda}} X_\mu X_\nu]$, and the set $\mathcal{X}$ contains Hermitian operators $X_\mu$ satisfying the local unbiased condition: $\Tr[\partial_\nu \rho_{\vec{\lambda}} X_\mu] = \delta_{\mu\nu}$. The Holevo bound becomes asymptotically achievable in the limit of collective measurements over many copies of the state\cite{hayashi2008,kahn2009,Yamagata2013}.
 \subsection{Incompatibility and Quantumness}
 A key challenge in multiparameter estimation arises from the incompatibility of optimal measurements for different parameters. The condition under which the SLD bound is saturable is given by the so-called weak commutativity criterion \cite{ragy2016}:
\begin{equation}
\Tr \left[\rho_{\vec{\lambda}} [L_\mu^S, L_\nu^S]\right]=0.\label{WCC}
\end{equation}
This motivates the definition of the antisymmetric matrix $U$, often called the mean Uhlmann curvature (MUC), capturing the average non-commutativity between parameter generators
\begin{equation}
{U}_{\mu\nu}:=\frac{1}{2i}\Tr \left[ \rho_{\vec{\lambda}} [L_\mu^S, L_\nu^S]\right]\,,
\end{equation}
and is useful to quantify the incompatibility between the pair of parameters 
$\lambda_\mu$ and $\lambda_\nu$.  
To quantify the extent to which quantum effects hinder classical-like estimation, the {\em quantumness} measure $R$  \cite{carollo2019, razavian2020quantumness} has been introduced
\begin{equation}
R:= ||iQ^{-1} U ||_{\infty},
\end{equation}
where $ ||\cdot||_{\infty}$ denotes the maximum eigenvalue of the matrix.  For the special case of two-parameter models, this simplifies to\cite{razavian2020quantumness}
\begin{equation}\label{defR}
R_2=\sqrt{\frac{\det\left[U\right]}{\det\left[ Q\right] }},
\end{equation}

For the case of three-parameter models, the quantumness measure admits a compact expression in terms of the quantum Fisher information matrix $Q$ and the elements of the Uhlmann matrix $U$. Specifically, we define the vector
\begin{equation}\label{u}
   \vec{u}=(U_{23},-U_{13},U_{12}), 
\end{equation}
and express quantumness as
	\begin{equation}\label{defR3}
	R_3=\sqrt{\frac{\vec{u}^T Q \vec{u}}{\det \left[ Q\right] }}.
	\end{equation}
This scalar quantity provides an operational quantifier of the estimation incompatibility arising from quantum mechanics, and serves as an upper bound on the relative gap between the Holevo and SLD bounds.
\section{Weight matrix-dependent measure}
Before introducing the weight matrix-dependent measure, we first briefly explain how the quantity $R$ is defined. Let us start with the expression of the Holevo bound
 \begin{equation*}
C_H[ W]
=\min_{ X \in \mathcal X} \Big\{\Tr\left[ W\, \textrm{Re}\left( Z[ X]\right)\right]+\Tr \left[\left|\left| W\, \textrm{Im}\left( Z[ X]\right)\right|\right|_1\right]\Big\},
\end{equation*}
 where $[Z]_{\mu\nu} := \Tr\left[\rho_{\vec{\lambda}} X_\mu X_\nu\right]$.
 When the operator $X_\mu$ in the set $\mathcal X$ is replaced by $\widetilde{X}_\mu := \sum_{\nu} [Q]_{\mu\nu} L_{\nu}$, the matrix $ Z$ in the Holevo bound is replaced by $\widetilde{ Z} := Q^{-1}(Q + iU)Q^{-1}$.
This leads to the following upper bound for the Holevo bound:
$$
C_H[W] \leq \Tr[WQ^{-1}] + \|\sqrt{W} Q^{-1} U Q^{-1} \sqrt{W}\|_1 ,
$$
further we have the inequality 
\begin{align}\label{inequality}
\|\sqrt{W} Q^{-1} U Q^{-1} \sqrt{W}\|_1 & \leq  \| Q^{-\half} W Q^{-1}UQ^{-\half}\|_1\leq \| Q^{-\half} W Q^{-\half}\|_1\|Q^{-\half}UQ^{-\half}\|_\infty
\\ \notag & =\Tr[WQ^{-1}] \cdot\|UQ^{-1}\|_\infty
\end{align}
Hence, we obtain
\begin{equation*}
C_H[W] - \Tr[WQ^{-1}] \leq \|\sqrt{W} Q^{-1} U Q^{-1} \sqrt{W}\|_1\leq \| Q^{-\half} W Q^{-1}UQ^{-\half}\|_1 \leq \Tr[WQ^{-1}] \cdot \|iU Q^{-1}\|_{\infty},
\end{equation*}
which implies the relative gap
\begin{equation*}
\frac{C_H[W] - \Tr[WQ^{-1}]}{\operatorname{Tr}[WQ^{-1}]} \leq \|iU Q^{-1}\|_{\infty}.
\end{equation*}
Quantumness is then defined as $R := \|iU Q^{-1}\|_{\infty}.$ 

This approach also allows us to define a potentially tighter bound via the normalized trace norm
\begin{equation*}
T[W] := \frac{\|\sqrt{W} Q^{-1} U Q^{-1} \sqrt{W}\|_1}{\operatorname{Tr}\left[WQ^{-1}\right]}.
\end{equation*}
leading to a hierarchy among different QCRBs incorporating $R$ and $T[W]$, as follows
\begin{align}
		C_{\SLD}[W]\leq C_H[W]\leq (1+T[W])C_{\SLD}[W]\leq (1+R)C_{\SLD}[W]\leq 2C_{\SLD}[W]\,.\label{ine}
\end{align}
We can naturally derive a bound for the quantity $T[W]$, namely,
\begin{equation*}
0 \leq T[W] \leq R.
\end{equation*}
The lower bound $T[W] = 0$ is achieved if and only if the SLD bound is attainable, i.e. when $U = 0$. Whether the upper bound $T[W] = R$ can be attained depends on the choice of the weight matrix $W$, and we analyze conditions for this attainability in the following section. 
For a diagonal weight matrix $W=\text{diag}(1,\omega)$, the $T$ bound is
\begin{equation}\label{eq:T2}
    T_2 = \frac{2 \sqrt{\text{det}[U]\omega}}{Q_{22}+\omega Q_{11}}.
\end{equation}
For the three-parameter model, the $T$ bound with a diagonal weight matrix $W = \mathrm{diag}(1,\omega_1,\omega_2)$ is given by
\begin{equation}
    T_3 = \frac{2\sqrt{\vec{u}^{T} Q \tilde{W}_3 Q \vec{u}}}{\sqrt{\det \left[ Q^2 \right]}},
\end{equation}
where $\vec{u}$ is defined in Eq.~(\ref{u}) and
\begin{equation*}
    \tilde{W}_3 = \mathrm{diag}(\omega_1\omega_2,\ \omega_2,\ \omega_1).
\end{equation*}
 The bound related to $T[W]$ (expressed as $C_T[W]=(1+T[W])C_{\SLD}[W]$), which lies between the Holevo bound $C_H[W]$ and $R$-dependent bound $C_R[W]=(1+R)C_{\SLD}[W]$, plays an important role. Studying its relation to these two bound is therefore essential. On one hand, although $C_H[W]$ has an operational meaning, it's difficult to express analytically and becomes hard to compute in cases involving more than two parameters. On the other hand, the $C_R[W]$ bound becomes increasingly less accurate as the dimension of estimation system grows, with the gap between it and $C_T[W]$ widening. As a result, we aim to further investigate the $C_T[W]$ bound to supply the limitations of these two bounds. Specifically, we explore this bound from two perspectives: its relationship with the Holevo bound $C_H[W]$, and its relationship with the $C_R[W]$ bound. 

\subsection{The Holevo bound $C_H[W]$ and the $T[W]$-related bound}
We begin by defining the geometric structure of the quantum statistical model.
\begin{definition}(SLD Tangent Space and Normal Space)
Consider an $n$-dimensional Hilbert space and a $d$-parameter family of quantum states $\rho_{\vec{\lambda}}$, where $n \geq d$. Let $L_i$ $(i=1,2,...,d)$ be the SLD operators at $\vec{\lambda}$, and their linear span is referred to as the SLD tangent space of the model at $\vec{\lambda}$, denoted as 
$$\mathcal{T}_{\vec{\lambda}}:=\text{span}_\mathbb{R}\{L_1,L_2,...,L_d\}.$$
Define the inner product as
$$
\langle L_{\nu},P_j\rangle _{\rho_{\vec{\lambda}}}= \frac{1}{2}\Tr[\rho_{\vec{\lambda}}\left(P_j L_{\nu}+L_{\nu}P_j\right)] = \Tr[\partial_\nu\rho_{\vec{\lambda}}P_j],
$$
where $P_j \in \mathcal{N}_{\vec{\lambda}}$ belongs to the normal space, satisfying $\Tr[\rho_{\vec{\lambda}}P_j]=0$, and the normal space $\mathcal{N}_{\vec{\lambda}}$ is orthogonal to the tangent space $\mathcal{T}_{\vec{\lambda}}$ under this inner product.
\end{definition}
The Holevo bound $C_H[W]$ is obtained from a global minimization over the set of all valid operators $\mathcal X$. The $T$-related bound $C_T[W]$ arises from a minimization restricted to the subspace $\mathcal{T}_{\vec{\lambda}}\cap \mathcal{X}$. Therefore, the equality $C_H[W]=C_T[W]$ holds if and only if a globally optimal operator can be found within this subspace. This means there must exist at least one optimal operator $X_\mu$ that lies entirely in the SLD tangent space. This insight motivates a formal decomposition of an arbitrary operator $X_\mu$ into its tangent and normal space components, which is provided by the following lemma.
\begin{lemma}
\label{CH}
Under the condition $n^2-1 > \text{dim}\mathcal{T}_{\vec{\lambda}}$, any operator $X_\mu \in \mathcal{X}$ satisfying $\Tr[\rho_{\vec{\lambda}}X_\mu]=0$ and $\Tr[\partial_\mu\rho_{\vec{\lambda}}X_\nu]=\delta_{\mu\nu}$ can be expressed as
$$
X_\mu = \sum_iL_i [Q^{-1}]_{i\mu}+ \sum_jP_j[K]_{j\mu},
$$
where $L_i \in \mathcal{T}_{\vec{\lambda}}$ are the SLD operators, and $P_j \in \mathcal{N}_{\vec{\lambda}}$ are elements of the normal space.
\end{lemma}
We generalize the Holevo bound formula from two- and three-parameter estimation 
\cite{Suzuki2016,bressanini2024multi} to $d$-parameter estimation in $n$-dimension quantum systems using Lemma~\ref{CH}.

\begin{theorem}\label{CH2}
The Holevo bound can be expressed as 
\begin{align}
C_H[W]=\min_{\substack{K \in \mathbb{R}^n} }&  \Big[
\Tr[WQ^{-1}]+\Tr[W \hbox{\rm Re}\left(K^TP_{\vec{\lambda}}K\right)] \notag \\ & +\|\sqrt W \left[Q^{-1}UQ^{-1}
+\hbox{\rm Im}\left(K^TP_{\vec{\lambda}}K\right) +2Q^{-1}SK \right]\sqrt W\|_1 \Big],
\end{align}
where $U$ is the Uhlmann matrix, and the dimensions of the matrices $K$, $S$, and $P_{\vec{\lambda}}$ are determined by the dimensions of the SLD tangent space and the normal space. 
Explicitly,
$$
[P_{\vec{\lambda}}]_{ij}=\Tr\!\big[\rho_{\vec{\lambda}} P_i P_j\big],
\qquad
[S]_{ij}=\hbox{\rm Im}\!\Tr\!\big[\rho_{\vec{\lambda}} L_i P_j\big].
$$
\end{theorem}
The theorem provides an alternative approach to evaluating $C_H[W]$. The proofs of the Lemma~\ref{CH} and Theorem~\ref{CH2} are given in the Appendices.

\subsection{The $T[W]$-related bound and the $R$-related bound}

We now examine the relation between $T[W]$ and $R$. We naturally derive a bound for the quantity $T[W]$, namely,
\begin{equation*}
0 \leq T[W] \leq R.
\end{equation*}
The lower bound $T[W] = 0$ is achieved if and only if the SLD bound is attainable, i.e. when $U = 0$. In contrast, the attainability of the upper bound depends on the choice of the weight matrix $W$. More specifically, $T[W]=R$ holds if and only if both inequalities in (\ref{inequality}) are satisfied simultaneously. In practical applications, diagonal $Q$ and $W$ are common, in which case these conditions can be analyzed explicitly for odd and even numbers of parameters.
\begin{theorem}
When $Q$ and $W$ are diagonal and the number of estimated parameters is odd, $T[W] = R$ implies $U = 0$. When even, $T[W] = R$ holds if $U = \bigoplus \begin{pmatrix} 0 & u \\ -u & 0 \end{pmatrix}$ with $u=\| Q^{-\half} UQ^{-\half}\|_\infty$.
\end{theorem}
    
The proofs are given in the Appendices. A simple sufficient condition for achieving $T[W] = R$ is that $W Q^{-1}$ is proportional to the identity, i.e. $W Q^{-1} = c I$ for some constant $c > 0$. To quantify how parameter incompatibility influences the gap between $T[W]$ and $R$, we establish the following theorem.
\begin{theorem}
\label{TR}
The gap between $T[W]$ and $R$ is higher when different parameters are incompatible:
$$T[W]\leq \text{Rank}(U)R.$$
\end{theorem}

This is easily proved by
\begin{align*} 
T[W]&=\frac{\|\sqrt{W} Q^{-1} U Q^{-1} \sqrt{W}\|_1}{\Tr[WQ^{-1}]}\leq\frac{\| Q^{-\half} W Q^{-1}UQ^{-\half}\|_1}{\Tr[WQ^{-1}]} \leq\frac{\| Q^{-\half} W Q^{-\half}\|_1\|Q^{-\half}UQ^{-\half}\|_1}{\Tr[WQ^{-1}]}\\
&\leq\sqrt{ \text{Rank}(U)}\|Q^{-\half}UQ^{-\half}\|_2\leq\sqrt{ \text{Rank}(U)}\|Q^{-\half}UQ^{-\half}\|_\infty=\text{Rank}(U)R,
\end{align*}
where $||\cdot||_2$ is the spectral norm.

\vspace{0.5em}

Theorem~\ref{TR} shows that the more incompatible the parameters, the larger the potential gap between $T[W]$ and $R$. In the 2-parameter qubit case, $\text{Rank}(U)=1$ and the bound is tight, while in higher dimensions the inequality quantifies how incompatibility limits simultaneous precision.

For a two-parameter estimation model, these results can be made explicit. Let $W = \text{diag}(1, \omega)$ be a diagonal weight matrix. Then
\[
T[W] = \frac{2\sqrt{\omega \det U}}{Q_{22} + \omega Q_{11}}.
\]
When $\det U \neq 0$ and $\det Q > 0$, we have $T[W] = R$ if and only if $Q_{12} = 0$. 
Moreover, $T[W]$ is maximized by choosing $\omega = Q_{22}/Q_{11}$.
For a non-diagonal weight matrix 
\[
W = \begin{pmatrix} 1 & \omega_1 \\ \omega_1 & \omega_2 \end{pmatrix},
\]
we have
\[
T[W] = \frac{2\sqrt{(\omega_2 - \omega_1^2)\det U}}{Q_{22} + \omega_2 Q_{11} - 2\omega_1 Q_{12}}.
\]
Then $T[W]=R$ holds if and only if 
\(\omega_2 = Q_{22}/Q_{11}\) and \(\omega_1 = Q_{12}/Q_{11}\).
The proof can be found in Appendix. Therefore, for 2-parameter  models, $R$ is equal to $T[W]$ for $W = \frac{1}{Q_{11}} Q$. In other words, when $W$ approaches $Q$, the value of $T[W]$ increases. For higher dimensional models, the calculation of $T$ becomes complex. We provide a more general expression for $T[W]$ in the Appendices. 
\section{Different forms of weight matrix and $T[W]$} 
The quantity $T[W]$ serves as a scalar performance measure in multiparameter quantum estimation. It depends on three elements: the positive definite weight matrix $W$, and the two matrices $Q$ and $U$, which are derived from the quantum states. In this section, we explore how different structures of $W$, specifically diagonal and non-diagonal forms, affect the expression and meaning of $T[W, Q, U]$.

When $W = I$, all parameters are considered equally important. In this case, $T[W, Q, U]$ reduces to the expression
\begin{align*}
T[I,Q,U] = \frac{\| Q^{-1} U Q^{-1} \|_1}{\Tr[Q^{-1}]},
\end{align*}
which serves as a measure of quantum incompatibility. This expression is invariant under reparameterization of the parameters.

Now consider a diagonal weight matrix $W = \mathrm{diag}(\omega_1, \dots, \omega_d)$, which anisotropically rescales the parameter space, assigning a relative importance $\omega_i$ to each parameter direction. This can be interpreted as adjusting the sensitivity or cost associated with estimation errors for each parameter.
We define transformed matrices
\begin{equation*}
Q'=\frac{1}{\sqrt{W}} Q\frac{1}{\sqrt{W}}, \;\;\;\;U'=\frac{1}{\sqrt{W}} U\frac{1}{\sqrt{W}},
\end{equation*}
and compute
\begin{align*}
T[I,Q',U']=\frac{||  Q'^{-1} U' Q'^{-1} ||_1}{\Tr[Q'^{-1}]}
=\frac{|| \sqrt{W} Q^{-1} U Q^{-1} \sqrt{W}  ||_1}{\Tr[WQ^{-1}]}=T[W,Q,U].
\end{align*}
This confirms that introducing a diagonal $W$ is equivalent to measuring $T[I,Q',U']$ in a rescaled parameter space. While $W$ acts only on the diagonal elements, its contribution to the trace norm arises solely through its interaction with $Q$. Since $U$ has zero diagonal entries (as it is the imaginary part of $\Tr[\rho_{\vec{\lambda}} L_i L_j]$), the diagonal $W$ does not affect $U$ directly. Nonetheless, for symmetry and geometric clarity, we retain the transformation of $U$ in the expression, emphasizing that $W$ represents a metric on the parameter space.

For a general (non-diagonal) positive definite weight matrix $W$, we write $W = P^\top D P$ via spectral decomposition, where $D = \mathrm{diag}(\omega_1, \dots, \omega_d)$ and $P \in \mathrm{SO}(d)$ is an orthogonal matrix. We define the rotated matrices as
\begin{equation*}
Q_r = P Q P^\top, \quad U_r = P U P^\top,
\end{equation*}
and compute
\begin{align*}
T[D,Q_r,U_r]=\frac{||  \sqrt{D}Q_r^{-1} U_rQ_r^{-1}  \sqrt{D}||_1}{\Tr[DQ_r^{-1}]}
=\frac{|| \sqrt{W}Q^{-1} UQ^{-1} \sqrt{W} ||_1}{\Tr[WQ^{-1}]}=T[W,Q,U],
\end{align*}
This shows that a general weight matrix $W$ corresponds to a rotation followed by anisotropic rescaling in parameter space. Physically, this amounts to expressing the estimation problem in a rotated coordinate system. Since quantum estimation bounds are invariant under orthogonal transformations (i.e. reparametrizations), the value of $T[W, Q, U]$ ultimately depends only on the relative geometry between $W$, $Q$, and $U$.
\section{Tunable 2-parameter qubit estimation model}
To demonstrate how the structure of $W$ affects estimation performance in concrete scenarios, we now introduce a tunable two-parameter qubit model where the SDL QCRB $C_{\SLD}$ is generally not attainable \cite{he2025}. This model allows for the comparison of different bounds, including $C_{H}[W]$, $C_T[W]$, and $C_R[W]$, under controllable quantum features such as state purity, parameter asymmetry, and 
unitary coupling.
The final state after sequential parameter encoding is given by
\begin{equation}
\rho_{\vec{\lambda}} = U_2 V U_1 \rho_0 U_1^\dagger V^\dagger U_2^\dagger,
\end{equation}
where the two parameters $\lambda_1$ and $\lambda_2$ are sequentially encoded onto the initial probe state $\rho_0$ via unitary operations. The parameter-encoding unitaries are defined as $U_k = e^{-i \sigma_3 \lambda_k}$ for $k = 1, 2$, corresponding to phase shifts generated by the Pauli-$Z$ operator.
To mitigate sloppiness and tune non-commutativity between the two parameter generators, i.e. 
to effectively modify the parameter-space geometry, we insert an intermediate unitary rotation
\begin{equation}
V = e^{-i \gamma \vec{\sigma} \cdot \vec{n}}, \quad \vec{n} = (\cos\phi \sin\theta,\, \sin\phi \sin\theta,\, \cos\theta),
\end{equation}
where $\vec{\sigma} = (\sigma_1, \sigma_2, \sigma_3)$ is the vector of Pauli matrices, and $\vec{n}$ is a unit vector specifying the rotation axis on the Bloch sphere. The parameter $\gamma$ controls the rotation angle. This intermediate rotation $V$ effectively mixes the two parameter directions, thereby introducing nontrivial quantum incompatibility and allowing for tunable coupling between the parameters.
We fix the weight matrix as $W = \mathrm{diag}(1, \omega)$ with $\omega > 0$, which is sufficient without loss of generality. This is because the intermediate unitary $V$ allows full control over the orientation of the parameter generators in Hilbert space, effectively enabling an arbitrary rotation of the coordinate axes in parameter space. In particular, the action of $V$ rotates the generators associated with each parameter, which is mathematically equivalent to diagonalizing any general (non-diagonal) weight matrix $W$ via a change of basis. 
Therefore, any symmetric positive definite $W$ can be transformed into diagonal form and absorbed into the model through an appropriate choice of $V$. This flexibility allows the model to simulate a wide range of cost structures, making it suitable for a systematic study of how quantum estimation bounds respond to variations in state purity, parameter asymmetry, and geometric structure of the weight matrix. 
 \subsection{Pure states}
\begin{figure}[htbp]
    \centering
    \includegraphics[width=0.6\textwidth]{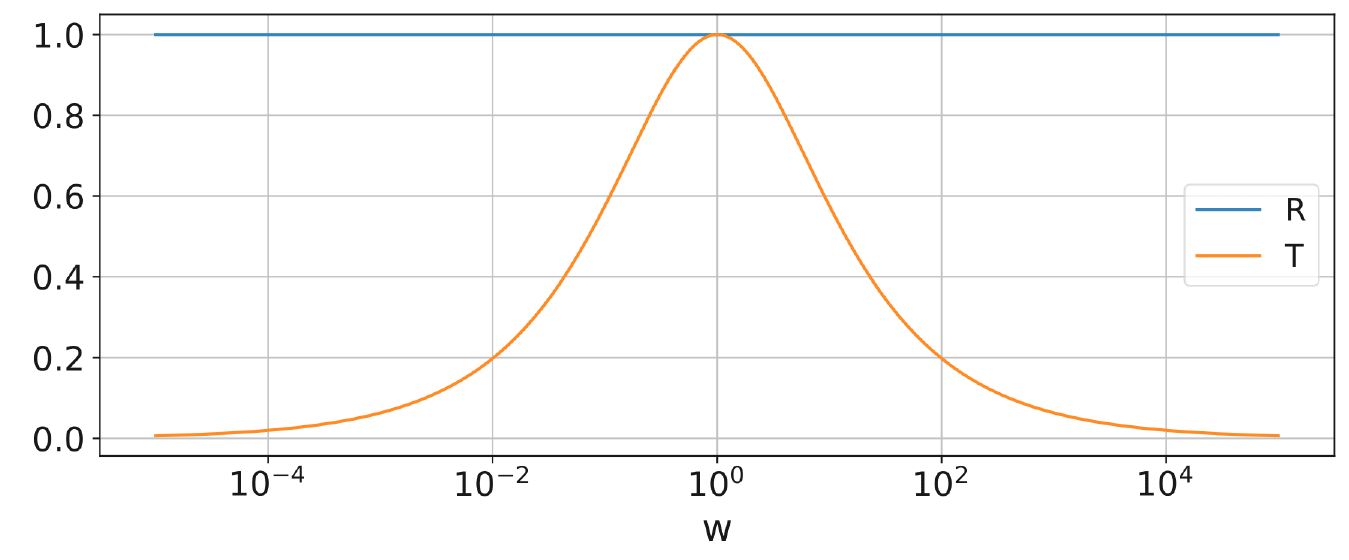}
    \caption{The two quantities $R$ and $T$ as a function of the weight asymmetry 
    parameter $\omega$ in a two-parameter model.}
    \label{fig:pure}
\end{figure}

For pure probe states of the form
\begin{equation*}
\ket{\psi_0}=\cos\frac{\alpha}{2} \ket{0}+ e^{i\beta}\sin  \frac{\alpha}{2}\ket{1}\,.
\end{equation*}
$C_H = C_T=\Tr[ W Q^{-1}]+ 2\sqrt{\det[ W Q^{-1}]}$ holds for all pure qubit states.
Fig.\ref{fig:pure} illustrates the behavior of $T[W]$ and $R$ as a function of the weight asymmetry parameter $\omega$, optimized over both pure states and intermediate rotations. This highlights the influence of cost structure on the attainability of estimation bounds even in the pure-state regime.

 \subsection{Mixed states}
We now consider mixed probe states of the form
\begin{equation*}
\rho_0 = \frac{1}{2}(I + \vec{r}_0 \cdot \vec{\sigma}),
\end{equation*}
where $\vec{r}_0 = (r_x, r_y, r_z)$ is the Bloch vector and $\vec{\sigma} = (\sigma_1, \sigma_2, \sigma_3)$ denotes the Pauli matrices. The two parameters $\lambda_1$ and $\lambda_2$ are sequentially encoded via $U_1$ and $U_2$. After encoding and intermediate rotation, the final state becomes
\begin{equation*}
\rho_{\vec{\lambda}} = U_2 V U_1 \rho_0 U_1^\dagger V^\dagger U_2^\dagger = \frac{1}{2}(I + \vec{r} \cdot \vec{\sigma}),
\end{equation*}
where $\vec{r} = R_z(2\lambda_2)R_{\vec{n}}(2\gamma)R_z(2\lambda_1)\vec{r}_0= (r_x', r_y', r_z')$ is the transformed Bloch vector, where $ R_z(\cdot)$ and $R_{\vec{n}}(\cdot)$ denote the rotation matrices about the 
$z$-axis and the axis $\vec{n}$, respectively. The expression of $\vec{r}$ and the explicit forms of the probe state, QFIM, and Uhlmann matrix elements are provided in Appendix. Without loss of generality, we set $\lambda_2 = 0$ to simplify the expressions. This is justified because $\lambda_2$ appears only through combinations with $\lambda_1$ and $\phi$, specifically in the forms $\xi=2\lambda_1 - \phi$ and $\varepsilon=2\lambda_2 + \phi$. Hence, fixing $\lambda_2 = 0$ is equivalent to a redefinition of angular variables and does not reduce the generality of our analysis.
 For the specific configuration $\gamma = \pi/4$ and $\theta = \pi/2$, which has been found to optimize metrological performance \cite{he2025}, the QFIM and Uhlmann matrix elements simplify to 
\begin{align}
Q_{11} &= 4(r_x^2 + r_y^2), \\
Q_{12} &= -4 r_z(r_y \cos\xi + r_x \sin \xi), \\
Q_{22} &= 2 \left[ |\vec{r}|^2+r_z^2 + (r_x^2 + r_y^2) \cos2\xi - 2 r_x r_y \sin 2\xi \right], \\
U_{12} &= 4 |\vec{r}|^2 ( -r_x \cos\xi + r_y \sin\xi),
\label{QU}
\end{align}
where
$$
|\vec{r}|^2 = r_x^2 + r_y^2 + r_z^2.
$$
These expressions yield the following fundamental identity
\begin{equation}
\label{eq:det_relation}
\det Q = |\vec{r}|^2 \det U,
\end{equation}
which connects the QFIM and the Uhlmann matrix through the purity of the quantum state. This relation generalizes the pure-state result $\det Q = \det U$ in \cite{he2025} to mixed states, incorporating the state's purity via the factor $|\vec{r}|^2$. Therefore, in the two-parameter mixed qubit model, we have $R=\frac{1}{|\vec{r}|}$.
The corresponding symmetric logarithmic derivative (SLD) operators take the form
\begin{align*}
L_{1} = \frac{\vec{y}_1 \cdot \vec{\sigma}}{1 - |\vec{r}|^2}, \;\;L_{2} = \frac{\vec{y}_2 \cdot \vec{\sigma}}{1 - |\vec{r}|^2},
\end{align*}
where the vectors $\vec{y}_1$ and $\vec{y}_2$ are given by
\begin{align*}
\vec{y}_1 &= \left\{ -2\cos\phi \, (r_y \cos\xi + r_x \sin \xi),\; -2\sin\phi \, (r_y \cos\xi + r_x \sin \xi),\; -2(r_y \sin\xi-r_x \cos\xi )\right\}, \\
\vec{y}_2 &= \left\{ 2[r_z\cos\phi + \sin\phi \, (r_y \sin\xi -r_x \cos\xi)],\; 2[r_z\sin\phi - \cos\phi \, (r_y \sin\xi-r_x \cos\xi)],\; 0 \right\}.
\end{align*}

\begin{figure}[htbp]
    \centering
    \includegraphics[width=0.9\textwidth]{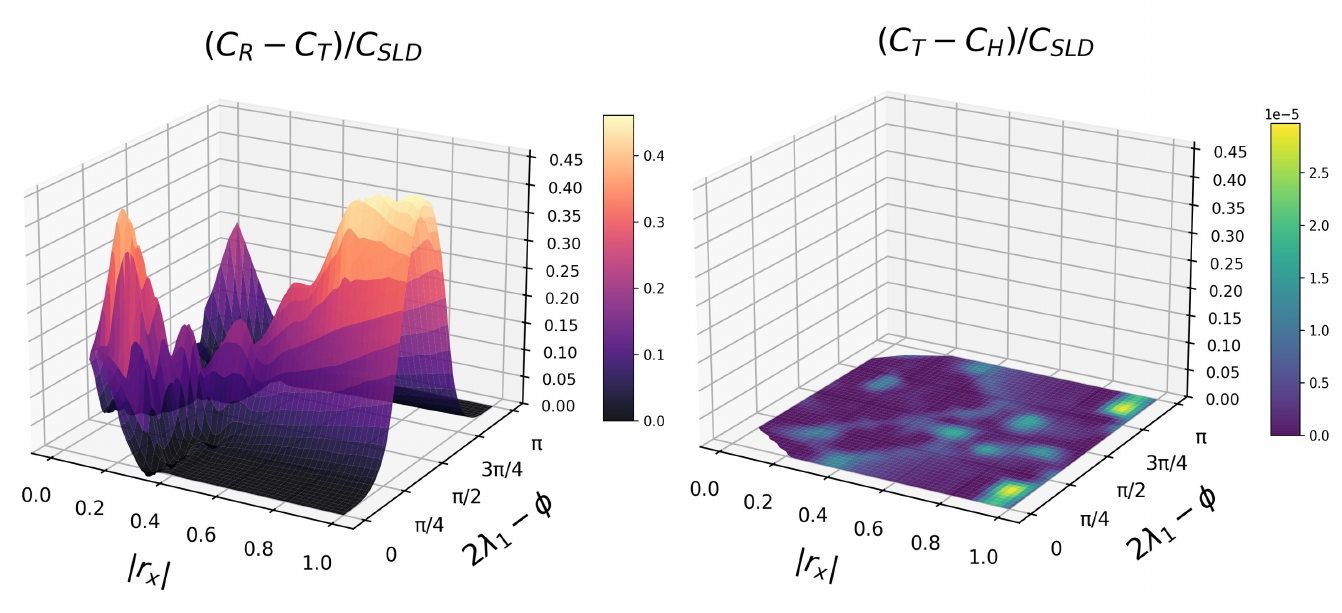}
    \caption{The ratios between the differences of bounds ($C_H$, $C_T$, and $C_R$) and the SLD bound are shown for mixed probe states under $\gamma = \pi/4$ and $\theta = \pi/2$. The difference between $C_H$ and $C_T$ remains consistently negligible across all configurations, whereas the gap between $R$ and $T$ exhibits a clear dependence on both $|r_x|$ and $\xi=2\lambda_1 - \phi$.}
    \label{fig:mixed_diff}
\end{figure}

Here we set $W=I$, so all bounds are expressed without reference to $W.$ Accordingly, $C_\SLD$ and $T$ are expressed as 
\begin{align*}
C_\SLD &= \frac{1}{4}\left(\frac1{ |\vec{r}|^2}+(r_y \sin\xi-r_x \cos\xi)^2\right), \\
T &= \frac{1}{2 |r_y \sin\xi-r_x \cos\xi|C_S }.
\end{align*}
We begin by comparing the bounds under optimal conditions ($\gamma = \pi/4$, $\theta = \pi/2$), and the estimation bounds $C_H$, $C_T$, and $C_R$ depend explicitly on the Bloch vector components $(r_x, r_y, r_z)$ and on the angular parameter $\xi$. In this study, the Holevo bound $C_H$ is evaluated using a semidefinite algorithm\cite{albarelli2019evaluating,genoni25}.
From Fig.~\ref{fig:mixed_diff}, the difference between $C_R$ and $C_T$ isn't directly related to the value of $|r_x|$ and it is much more susceptible to the influence of $\xi$. 
$(C_R-C_T)/C_\SLD$ shows a pronounced enhancement at $\xi = \pi/2$. When $|r_x|$ approaches 1, the value of $R$ converges to 1, and consequently, the value of $T$ strongly dependents on the optimization of the system. In contrast, the difference between $C_T$ and $C_H$ remains negligible, except at singular parameter values where yellow spots appear in the figure. These singularities, however, do not affect our conclusion that $C_T$ serves as a reliable substitute for $C_R$ in two-parameter mixed-state scenarios. 
 \begin{figure}[htbp]\label{diffplot2}
    \centering
    \begin{subfigure}[b]{0.53\textwidth}
        \centering
        \includegraphics[width=\textwidth]{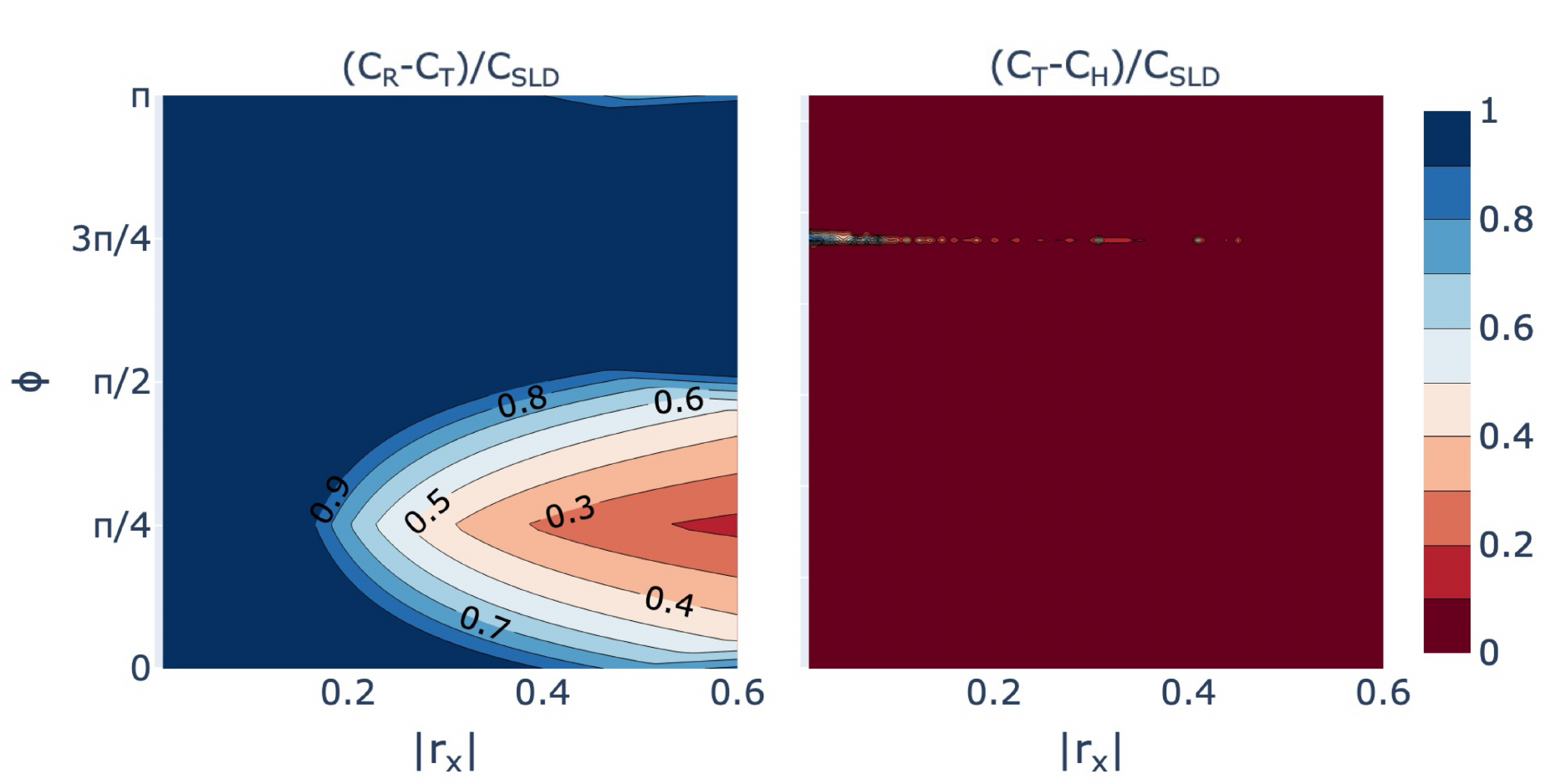}
        \caption{$|r_x|$ vs. $\phi$ when $r_x = r_y$, $r_z = 0.5$}
        \label{fig:subfig_a}
    \end{subfigure}
    \hspace{-2.8em}
    \begin{subfigure}[b]{0.52\textwidth}
        \centering
        \includegraphics[width=\textwidth]{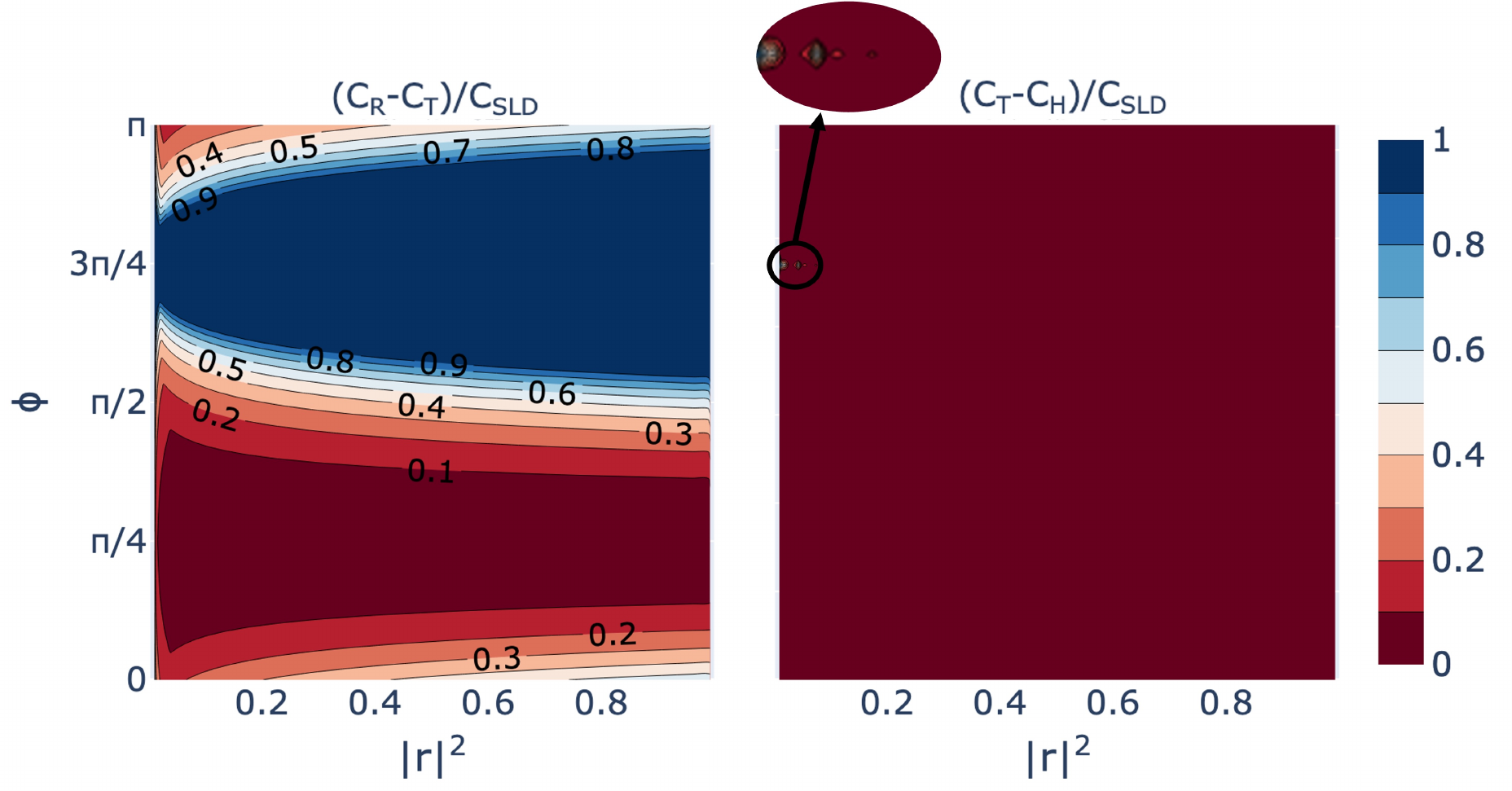}
        \caption{$|\vec{r}|^2$ vs. $\phi$ when $r_x = r_y$, $r_z = 0.1$}
        \label{fig:subfig_b}
    \end{subfigure}
    \caption{The ratios between the differences of bounds ($C_H$, $C_T$, and $C_R$) and the SLD bound under fixed encoding ($\lambda_1 = \lambda_2 = 0$) and $\gamma = \pi/4$, $\theta = \pi/2$, showing their dependence on (a) $|r_x|$ and (b) $|\vec{r}|^2$ as $\phi$ varies.}
    \label{fig:diff2}
\end{figure}
As shown in SLD operators' expressions, when $r_y \sin \xi - r_x \cos \xi \neq 0$, $\vec{y}_1$ and $\vec{y}_2$ are linearly independent, so the SLD tangent space is two-dimensional, matching the number of parameters, with one remaining normal direction. Assuming the existence of a matrix $P=\vec{p}\cdot\vec{\sigma}$ in the normal space, the conditions $\mathrm{Tr}[\rho_{\vec{\lambda}}P]=0$ and $\mathrm{Tr}[\rho_{\vec{\lambda}}P(PL_i+L_iP)]=0$ respectively imply $\vec{r}\cdot\vec{p}=0$ and $\vec{p}\propto\vec{y}_1\times\vec{y}_2$. This leads to $\vec{r}\cdot(\vec{y}_1\times\vec{y}_2)=0$, suggesting $\vec{r}$ lies in the plane spanned by $\vec{y}_1$ and $\vec{y}_2$. Yet, $\mathrm{Tr}[\rho_{\vec{\lambda}}L_i]=0$ requires $\vec{r}$ to be orthogonal to both. These conditions are contradictory, and thus no such matrix exists in the normal space. This is also consistent with what we observe in the Fig.~\ref{fig:mixed_diff}, where $C_H$ is always equal to $C_T$.
When $r_y \sin \xi - r_x \cos \xi = 0$, $\vec{y}_1$ and $\vec{y}_2$ are linearly dependent. From Eqs.~(\ref{QU})  and (\ref{eq:det_relation}) we can esaily find the the QFIM become singular. We examine some special points with $\lambda_1 = \lambda_2 = 0$, $\gamma = \pi/4$, $\theta = \pi/2$, $r_x = r_y$, and fixed $r_z$ to see how different bounds perform. As shown in Fig.~\ref{fig:diff2}, when $\phi = 3\pi/4$, the two SLD operators become linearly dependent and the tangent space is one-dimensional, resulting in a larger difference between $C_H$ and $C_T$. Moreover, as $r_z$ declines, the extent of this elongated region decreases. In contrast, the difference between $C_R$ and $C_T$ does not clearly reveal the singularity of the QFIM. Their gap persists across parameter space, and only by reducing the mixedness can the difference in certain regions be noticeably reduced.
This indicates that a larger tangent space is beneficial for extracting information about both parameters. However, in the two-parameter mixed qubit model, a two-dimensional tangent space already provides the optimal setting for improving the accuracy of parameter estimation.

The above observations suggest that $C_H$ and $C_T$ coincide under the configurations considered. To assess the generality of this equality and to explore possible deviations, we next investigate a different class of probe states and parameter settings.
 \section{Multiparameter estimation for SU(2) unitary encoding}
Unitary parameterization constitutes an important and widely used class of models in quantum metrology. Such parameterizations naturally arise in tasks like phase estimation in optical interferometry and magnetometry, where physical parameters are encoded in the state through unitary dynamics (see also \cite{pal2025role}). The unitary parameterization process can be expressed as
$$
\rho(\vec{\lambda}) = U(\vec{\lambda}) \rho_0 U^\dagger(\vec{\lambda}), 
$$
where $\rho_0$  is an initial state independent of the parameter vector $ \vec{\lambda}$, and $U(\vec{\lambda})$ is a unitary operator satisfying
$$U(\vec{\lambda})  U^\dagger(\vec{\lambda})  = U^\dagger(\vec{\lambda}) U(\vec{\lambda}) = I.$$ 
In this Section, we mainly consider the case where $ \rho_0$ is a pure state, i.e. $ \rho_0=\ket{\psi_0}\bra{\psi_0}$ and notice that for unitary parameterization acting on a fixed initial state, the structure of the SLD becomes closely related to the generators of the unitary evolution. This observation motivates introducing the Hermitian operator \cite{liu2015quantum}
$$
\mathcal{H}_k = i \left( \partial_k U \right) U^\dagger,
$$
which acts as the generator of translations along the parameter $\lambda_k$ in the unitary group. For pure states, the QFIM and the Uhlmann matrix have a straightforward form in terms of these generators:
\begin{align*}
Q_{kl}&=2 \langle\psi_0|\{ \mathcal{H}_k,\mathcal{H}_l\}|\psi_0\rangle - 4\langle\psi_0| \mathcal{H}_k|\psi_0\rangle  \langle\psi_0| \mathcal{H}_l|\psi_0\rangle,\\
U_{kl}&=-2i \langle\psi_0|[ \mathcal{H}_k,\mathcal{H}_l]|\psi_0\rangle.
\end{align*}
When the unitary evolution is generated by a Hamiltonian $H_{\vec{\lambda}}$ as $U = e^{-i t H_{\vec{\lambda}}}$, $\mathcal{H}_{k} $ can be expressed as an infinite series involving nested commutators \cite{liu2015quantum}:
\begin{equation*}
\mathcal{H}_{k} = i \sum_{n=0}^{+\infty} f_n H_{\vec{\lambda}}^{\times} (\partial_{k} H_{\vec{\lambda}})^{n}, \,\,\,\,\,f_n = \frac{(i t)^{n+1}}{(n+1)!}, 
\end{equation*}
with the superoperator $H_{\vec{\lambda}}^{\times} (A) = [H_{\vec{\lambda}}, A]$. This series often admits a closed form for $SU(2)$ models, which we exploit in the following example.
 
Let us now consider the Hamiltonian\cite{Candeloro2024}
\begin{equation*}
    H_{B,\theta} = B\left[\cos(\theta) J_x + \sin(\theta) J_z\right] \,,
\end{equation*}
with $J_k$ the $k$-th generator of $SU(2)$ and $J_{\vec{n}_\theta}=\vec{n}_\theta\cdot \vec{J}$ with $\vec{J}=(J_x, J_y, J_z)$  satisfying the commutation relations
\begin{align*}
    [J_j, J_k] &= i \epsilon_{jkl}J_l. 
\end{align*}
The calculation of QFIM and Uhlmann matrix are shown in Appendix. 
\subsection{2-parameter estimation for qubit}
\begin{figure}[htbp]
      \centering
        \includegraphics[width=0.8\textwidth]{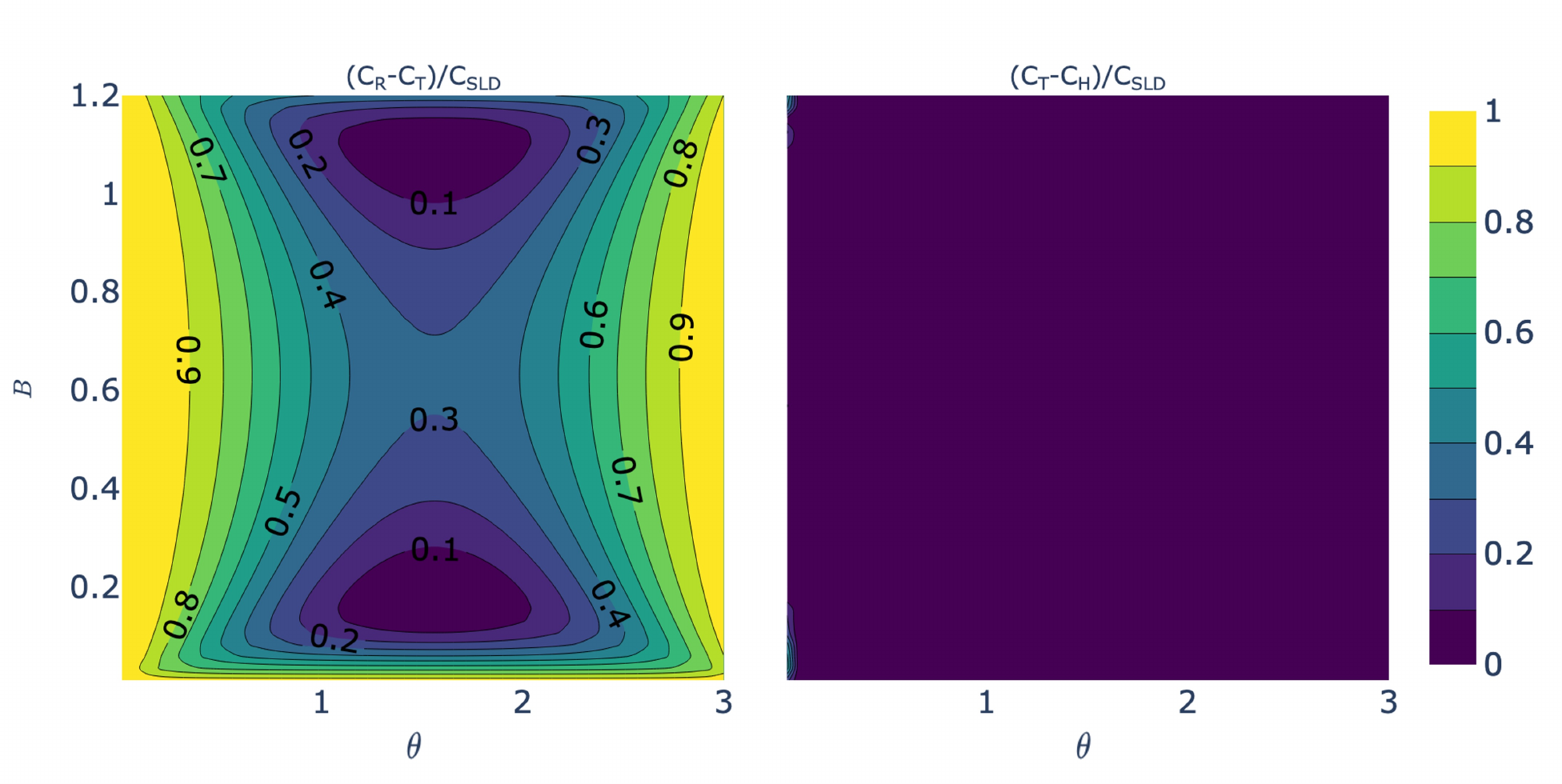}
    \caption{The ratios between the differences of bounds ($C_H$, $C_T$, and $C_R$) and the SLD bound are shown as functions of $\theta$ and $B$ when $\alpha=\pi/2$ and $t=5$.}
    \label{fig3}
\end{figure}

For 2-parameter estimation qubit model, we set the pure initial state
$$
\ket{\psi_0} =\cos\frac{\alpha}{2} \ket{1} + e^{i\beta} \sin\frac{\alpha}{2} \ket{0}.
$$
The SLD operators are given by
\begin{align*}
   L_\theta&=4\sin\frac{Bt}{2}(\vec{n}_1\times \vec{r}_0) \cdot\vec{J},\\
    L_B&=-2(\vec{n}_\theta\times \vec{r}_0)\cdot \vec{J},
\end{align*}
with
\begin{align*}
    &\vec{n}_\theta = (\cos \theta, 0, \sin \theta),\\
    &\vec{n}_1 = (\cos\frac{Bt}{2}\sin\theta,-\sin\frac{Bt}{2},-\cos\frac{Bt}{2}\cos\theta),\\
    &\vec{r}_0 = (\sin\alpha\cos\beta,\sin\alpha\sin\beta, \cos\alpha).
\end{align*}
When we fix $\alpha$ in initial state and $t=5$, we can investigate how the differences between $C_H$, $C_T$, and $C_R$. For example, when $\alpha=\pi/2$,  the dimension of the SLD operator space depends on the values of parameters. The differences between $C_H$, $C_T$, and $C_R$ are shown in Fig.\ref{fig3} with $B$ and $\theta$ changing. $C_R$ and $C_T$ are always not equal. But $C_T$ are always close to $C_H$.   

\subsection{3-parameter estimation for qutrit}
Having studied the two-parameter case in the previous section, we now consider the compatibility of estimating three parameters in an $SU(2)$ unitary encoding. The generator takes the form\cite{Candeloro2024}
\begin{equation}
    H_{B, \theta, \varphi} = H = B J_{\vec{n}^{(3)}_\theta}
    \tag{52}
\end{equation}
where $J_x, J_y, J_z$ are elements of the $SU(2)$ algebra. We use the notation
$$
J^{(3)}_{\vec{n}_\theta} = \vec{n}_\theta \cdot \vec{J}
$$
with the direction vector
$$
\vec{n}_\theta^{(3)} = (\cos\theta \cos\varphi, \cos\theta \sin\varphi, \sin\theta)
$$
and $\vec{J} = (J_x, J_y, J_z)$.
We consider the initial state 
\begin{equation*}
\ket{\psi_0} =\cos\frac{\alpha}{2} \ket{1} + e^{i\beta} \sin\frac{\alpha}{2} \ket{-1}.
\end{equation*}
 The QFIM and Uhlimann  matrix are provided in Appendix. We find
\begin{align*}
\text{det}Q&=64\, t^{2}\cos^{2}\theta \,\sin^{4}\!\frac{Bt}{2}\,\sin^{2}2\alpha,\\
\text{det}U&=0.
\end{align*}
Unlike the 2-parameter qubit estimation model, where $\text{det}Q=|\vec{r}|^2\text{det}U$ always holds, this equality does not generally hold in the three-parameter qutrit estimation model. Moreover, it does not hold for all $2\times2$ submatrices either. The SLD operators are shown in Appendix. 

We fix the initial state and the values of the parameters to apply Theorem \ref{CH2} in order to compute the Holevo bound. Specifically, we set $\alpha=\pi/4$, $t=1,$  $B=\pi,$  $\theta=0,$  $\varphi=0$ and find that the Holevo bound is achieved optimal with 
$C_H=C_T=\frac{11+\sqrt2}{8}\approx1.5518$ with $K=\bf{0}$. The detailed calculation is presented in the Appendix.

 \begin{figure}[htbp]
      \centering     \includegraphics[width=0.8\textwidth]{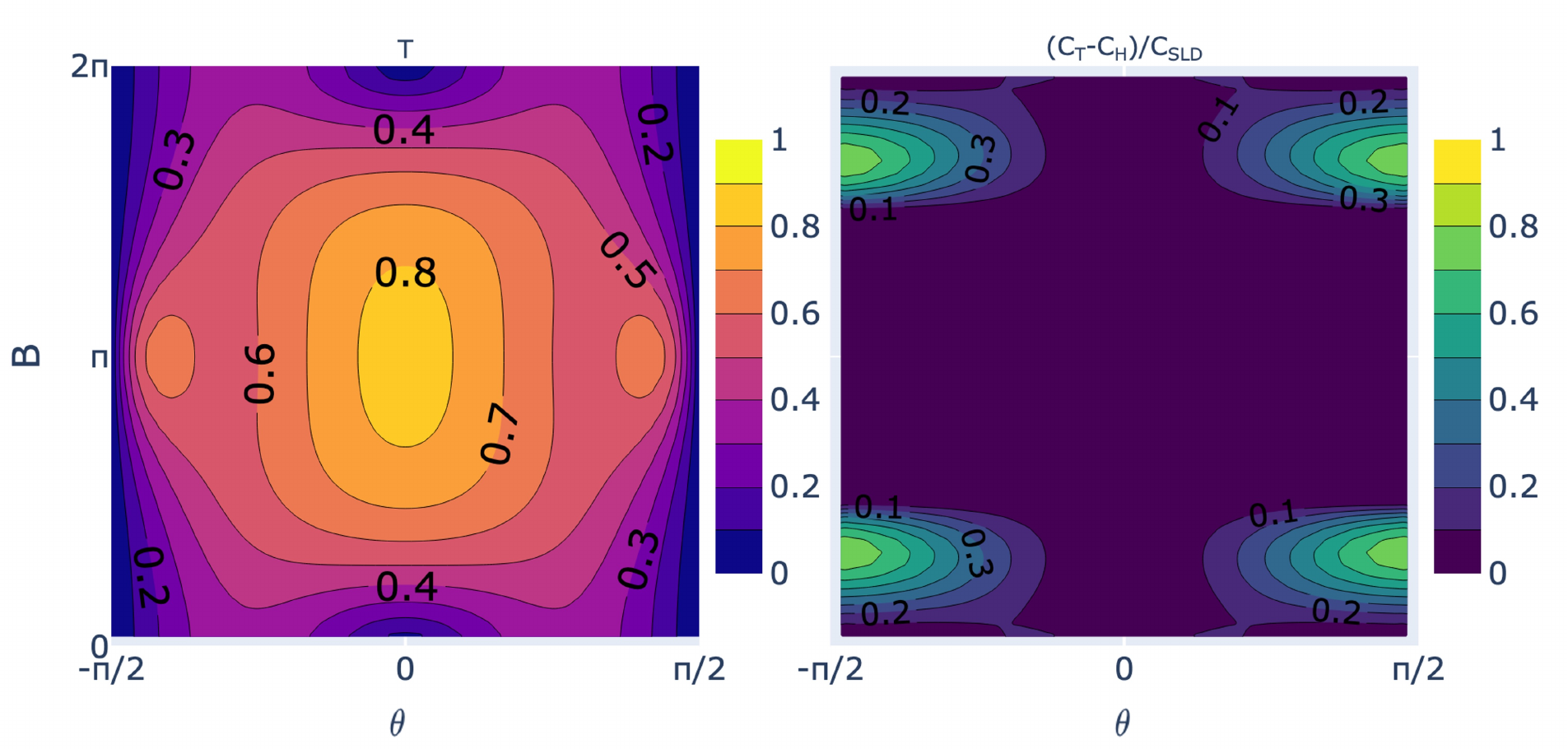}
    \caption{$T$ and the ratio between $C_T-C_H$ and $C_\SLD$ are shown as functions of $\theta$ and $B$ when $\alpha=\pi/4$, $\beta=0$, $\varphi=0$ and $t=1$.}
    \label{3parqutrit}
\end{figure}

Since $R=1$ always holds in the three-parameter pure qutrit case, our focus is on the difference between $C_H$ and $C_T$. As shown in Fig.~\ref{3parqutrit}, the gap between $C_T$ and $C_H$ is not always negligible. However, as the parameters approach the central point, the value of $T$ increases while the gap remains nearly zero, which is consistent with our previous analytical results that  $C_T=C_H$ when $B=\pi$ and $\theta=0$.
\section{Conclusion and discussion}
\label{sec:outro}
In this work, we have presented a detailed investigation about the role of the weight matrix $W$ in quantifying quantum incompatibility in multiparameter estimation. We have clarified the relationship between different quantum Cram\'er-Rao bounds by discussing the role of a weight-dependent measure $T[W]$ and comparing it to the weight-independent quantumness measure $R$. In particular, we have analyzed in details  the  hierarchy of bounds: $C_{\SLD}[W] \leq C_H[W] \leq C_T[W] \leq C_R[W] \leq 2C_{\SLD}[W]$, where $C_T[W] = (1+T[W])C_{\SLD}[W]$ and $C_R[W] = (1+R)C_{\SLD}[W]$, proving that the gap between the SLD bound $C_\SLD[W]$ and the Holevo bound $C_H[W]$ is not a fixed property of the quantum statistical model alone but is modulated by the choice of $W$, which defines the weight function for estimation precision. The measure $T(W) = \|\sqrt{W}Q^{-1}UQ^{-1}\sqrt{W}\|_{1} / \mathrm{Tr}[WQ^{-1}]$ provides a more refined and often tighter quantification of this gap than $R$. We have 
derived explicit conditions under which the bounds $C_H[W]$ and $C_T[W]$ coincide, linking this to the structure of the optimal measurement operators in the SLD tangent space.

Through analytical and numerical studies of concrete models, we have 
validated these theoretical findings. For two-parameter qubit models, we have 
confirmed that $C_H[W] = C_T[W]$ holds generally, while the gap between $C_T[W]$ and $C_R[W]$ varies significantly with the Bloch vector of the state and  the weight asymmetry parameter $\omega$. We have also shown that $T[W]$ reaches its maximum, equal to 
$R$, when the weight matrix is proportional to the QFIM, $W \propto Q$. The analysis of 
$SU(2)$  unitary encoding models reinforced these results. For the two-parameter qubit case, $C_T[W]$ again proved to be an excellent approximation to $C_H[W]$ and, more significantly, in the three-parameter qutrit model, we find that  $C_H$ and $C_T$ coincide in certain parameter regimes (see also \cite{pal2025role}), showing that their equality can still persist beyond the qubit case. Even when the equality $C_H = C_T$ does not hold, Theorem~\ref{CH2} provides an analytical expression for $C_H$ in terms of $C_T$, which enables us to compute the Holevo bound efficiently. A result emerging from our analysis is the identity $\det Q = |\vec{r}|^2 \det U$ for the two-parameter mixed qubit model, which generalizes the pure-state relation by incorporating the state purity and shows 
that mixedness can mitigate the effects of quantum incompatibility.

Our results opens promising avenues for future research, including the generalization to arbitrary dimensions to understand how the dimensionality of the SLD tangent space and its complement (the normal space) influences the gaps. More importantly,  our results emphasize that the hardness of an estimation task depends on the interplay between the quantum model $(Q, U)$ and the cost function $W$. This suggests a new paradigm for optimal experiment design: finding probe states and (possibly adaptive) measurement schemes that minimize not the SLD bound $C_{\SLD}$, but the more relevant incompatibility-aware bounds $C_T$ or $C_H$ for a given $W$. 

In conclusion, moving beyond the weight-independent quantumness $R$ to the weight-dependent measure $T(W)$ provides a more accurate and practical tool for assessing the limits of multiparameter quantum estimation. By explicitly incorporating the cost of errors, this approach brings the theory closer to real-world applications, where not all parameters are created equal.
\section*{Appendix}
\subsection{Proof of Lemma III.1}
Let the space of relevant Hermitian operators be a Hilbert space equipped with the inner product $\langle A,B\rangle_{\rho_{\vec{\lambda}}}$ defined previously. This space can be decomposed into the direct sum of the SLD tangent space and its orthogonal complement, the normal space: $\mathcal{H}=\mathcal{T}_{\vec{\lambda}}\oplus\mathcal{N}_{\vec{\lambda}}.$ Consequently, any operator 
$X_{\mu}\in \mathcal{X}$ admits a unique decomposition into a component in the tangent space and a component in the normal space: $$X_\mu=Y_\mu+V_\mu,$$
where $Y_{\mu}\in\mathcal{T}_{\vec{\lambda}}$ and $V_{\mu}\in \mathcal{N}_{\vec{\lambda}}$.

By definition of the tangent space, $Y_{\mu}$ can be expressed as a linear combination of the SLD basis operators, $Y_{\mu}=\sum_{i=1}^d c_{i\mu }L_i$ with $c_{i\mu }\in\mathbb{R}$. Similarly, $V_{\mu}=\sum_jP_j[K]_{j\mu}$ for some real matrices $K$. 

The unbiased estimator condition for $X_{\mu}$ is $\langle L_{\nu},X_{\mu}\rangle_{\rho_{\vec{\lambda}}}=\delta_{\mu\nu}$. Substituting the decomposition, we get
$$\langle L_{\nu},X_{\mu}\rangle_{\rho_{\vec{\lambda}}}=\langle L_{\nu},Y_{\mu}\rangle_{\rho_{\vec{\lambda}}}+\langle L_{\nu},V_{\mu}\rangle_{\rho_{\vec{\lambda}}}=\delta_{\mu\nu}.$$
By the definition of the normal space, its elements are orthogonal to the tangent space. Using the definition of the inner product, the orthogonality condition $\langle L_{\nu},V_{\mu}\rangle_{\rho_{\vec{\lambda}}}=0$ directly implies that
$\langle L_{\nu},Y_{\mu}\rangle_{\rho_{\vec{\lambda}}}=\delta_{\mu\nu}.$
We now substitute the expansion of $Y_{\mu}$ to solve for the coefficients $c_{i\mu }$:
$$\langle L_{\nu},Y_{\mu}\rangle_{\rho_{\vec{\lambda}}}=\sum_{i=1}^d c_{i \mu }\Tr[\partial_\nu\rho_{\vec{\lambda}}L_{i}]=\sum_{i=1}^d [Q]_{\nu i}c_{i \mu } =\delta_{\mu\nu}.$$
This is a matrix equation. Let $C$
be the matrix with elements $[C]_{i\mu}=c_{i \mu }$. The equation can be written in matrix form as $QC=I.$ Assuming the QFIM is invertible, we can solve for 
$C$ by left-multiplying by $Q^{-1}$:
$$C=Q^{-1}.$$
Therefore, the coefficients are given by $c_{i\mu}=[Q^{-1}]_{i \mu},$ and substituting this back into the $Y_{\mu}$ yields $$Y_{\mu}=\sum_{i=1}^d L_i[Q^{-1}]_{i \mu}.$$

Combining this with the decomposition $X_\mu=Y_\mu+V_\mu$ completes the proof.
\\

\subsection{Proof of Theorem III.2}
From Lemma $\ref{CH}$ we know 
$$X_\mu=Y_\mu+V_\mu=\sum_iL_i[Q^{-1}]_{i\mu}+\sum_jP_j[K]_{j\mu}.$$
Because $V_\mu\in  \mathcal{N}_{\vec{\lambda}}$ we have
\begin{align*}
Z_{\mu\nu}&=\Tr[\rho_{\vec{\lambda}}X_\mu X_\nu]=\Tr[\rho_{\vec{\lambda}}Y_\mu Y_\nu]+\Tr[\rho_{\vec{\lambda}}V_\mu V_\nu]+\Tr[\rho_{\vec{\lambda}}Y_\mu V_\nu]+\Tr[\rho_{\vec{\lambda}}V_\mu Y_\nu]\\
&=\sum_{i,j}[Q^{-1}]_{i\mu}\Tr[\rho_{\vec{\lambda}}L_iL_j][Q^{-1}]_{j\nu}+\sum_{i,j}[K]_{i\mu}\Tr[\rho_{\vec{\lambda}}P_iP_j][K]_{j\nu}\\
&\;\;+\sum_{i,j}[Q^{-1}]_{i\mu}\Tr[\rho_{\vec{\lambda}}L_iP_j][K]_{j\nu}+\sum_{i,j}[Q^{-1}]_{i\nu}\Tr[\rho_{\vec{\lambda}}P_iL_j][K]_{j\mu},
\end{align*}

we define $[L_{\vec{\lambda}}]_{i,j}=\Tr[\rho_{\vec{\lambda}}L_iL_j],$ $[P_{\vec{\lambda}}]_{i,j}=\Tr[\rho_{\vec{\lambda}}P_iP_j],$ and $[S]_{i,j}=\text{Im}\Tr[\rho_{\vec{\lambda}}L_iP_j]$. We obtain
\begin{align*}
Z_{\mu\nu}&=[Q^{-1}L_{\vec{\lambda}}Q^{-1}]_{\mu\nu}+[K^TP_{\vec{\lambda}}K]_{\mu\nu}+[Q^{-1}SK]_{\mu\nu}+[K^TSQ^{-1}]_{\mu\nu}\\
&=[Q^{-1}L_{\vec{\lambda}}Q^{-1}]_{\mu\nu}+[K^TP_{\vec{\lambda}}K]_{\mu\nu}+[Q^{-1}SK]_{\mu\nu}-[Q^{-1}SK]_{\nu\mu}\\
&=[Q^{-1}L_{\vec{\lambda}}Q^{-1}]_{\mu\nu}+[K^TP_{\vec{\lambda}}K]_{\mu\nu}+2[Q^{-1}SK]_{\mu\nu},
\end{align*}
 and  
\begin{align*}
h[Z|W]&=\Tr[W\text{Re} Z]+\|\sqrt W\text{Im} Z\sqrt W\|_1\\
&=\Tr[W\text{Re} \left(Q^{-1}L_{\vec{\lambda}}Q^{-1}+K^TP_{\vec{\lambda}}K\right)]+\|\sqrt W\text{Im} \left(Q^{-1}L_{\vec{\lambda}}Q^{-1}+K^TP_{\vec{\lambda}}K+2Q^{-1}SK\right)\sqrt W\|_1\\
&=\Tr[WQ^{-1}]+\Tr[W\text{Re}\left(K^TP_{\vec{\lambda}}K\right )]+\|\sqrt W(Q^{-1}UQ^{-1}+\text{Im}\left (K^TP_{\vec{\lambda}}K\right )+2Q^{-1}SK)\sqrt W\|_1,
\end{align*}
and 
\begin{equation*}
C_H[W]=\min_{\substack{K \in \mathbb{R}^n} }\left[\Tr[WQ^{-1}]+\Tr[W\text{Re}\left (K^TP_{\vec{\lambda}}K\right )]+\|\sqrt W(Q^{-1}UQ^{-1}+\text{Im}\left (K^TP_{\vec{\lambda}}K\right)+2Q^{-1}SK)\sqrt W\|_1\right].
\end{equation*}
\\
\subsection{Proof of Theorem III.3}
First, we prove that the second inequality in Eq.~(\ref{inequality})
$$ \| Q^{-\half} W Q^{-1}UQ^{-\half}\|_1= \| Q^{-\half} W Q^{-\half}\|_1\|Q^{-\half}UQ^{-\half}\|_\infty$$ 
holds if and only if there exists an eigenvalue decomposition 
$$Q^{-\half} W Q^{-\half}=\sum_{i=1}^n\sigma_i\op{v_i}{v_i}$$
such that the vectors $\left\{ (Q^{-\half} U Q^{-\half})^* \ket{v_i} \right\}_{i=1}^n$ are pairwise orthogonal and satisfy 
$$\left\| (Q^{-\half} U Q^{-\half})^* \ket{v_i} \right\|_2 = \left\| Q^{-\half} U Q^{-\half} \right\|_\infty \quad \text{for} \;\;\;\;\forall i.$$

To prove this, we consider the eigenvalue decomposition $Q^{-\half} W Q^{-\half}=\sum_{i=1}^n\sigma_i\op{v_i}{v_i}$, we have 
\begin{align*}
\left\| Q^{-\half} W Q^{-1} U Q^{-\half} \right\|_1 = \left\| \left( \sum_{i=1}^n \sigma_i \op{v_i}{v_i} \right) (Q^{-\half} U Q^{-\half}) \right\|_1.
\end{align*}

Define $\ket{\phi_i} := (Q^{-\half} U Q^{-\half})^* \ket{v_i}$. The right-hand side is
$$
\left\| Q^{-\half} W Q^{-\half} \right\|_1 \left\| Q^{-\half} U Q^{-\half} \right\|_\infty = \left( \sum_{i=1}^n \sigma_i \right) \left\| Q^{-\half} U Q^{-\half} \right\|_\infty.
$$
By the trace norm equality condition for products of operators, equality holds if and only if $\{\ket{\phi_j}\}$ are orthogonal and satisfy $\| \ket{\phi_i} \|_2 = \left\| Q^{-\half} U Q^{-\half} \right\|_\infty$ for all $i$.

Then consider the eigenvalue decomposition $Q=\sum_{i=1}^n q_i\op{i}{i}$ and $W=\sum_{i=1}^n w_i\op{i}{i}$, for the first inequality in Eq.~(\ref{inequality}), we have 
\begin{align*}
\sqrt{W} Q^{-1} U Q^{-1} \sqrt{W}= \sum_{i=1}^n \frac{\sqrt{w_i }}{q_i}\op{i}{i}U \sum_{j=1}^n \frac{\sqrt{w_j }}{q_j}\op{j}{j}=\sum_{i,j=1}^n \frac{\sqrt{w_i w_j}}{q_iq_j}\langle i | U | j \rangle\op{i}{j}.
\end{align*}

The right-hand side is
$$
Q^{-\half} W Q^{-1} U Q^{-\half} =\sum_{i=1}^n \frac{w_i}{\sqrt{q_i }q_i}\op{i}{i}U \sum_{j=1}^n \frac{1}{\sqrt{q_j}}\op{j}{j}=\sum_{i,j=1}^n\frac{w_i}{\sqrt{q_i q_j}q_i}\langle i | U | j \rangle\op{i}{j}.
$$

Therefore, the left-hand side and the right-hand side are equal if and only if
$$
\frac{w_i}{q_i}=c\;\;\; \forall i,
$$
where $c\in\mathbb{R}^+$. This condition means $W=cQ$.

Under this condition, the second inequality in Eq.~$(\ref{inequality})$ becomes $$\| Q^{-\half} UQ^{-\half}\|_1= \| Q^{-\half} UQ^{-\half}\|_\infty,$$ which holds only and if only all singular values of $ Q^{-\half} UQ^{-\half}$ are equal. 
 Thus, when the number of independent parameters is odd, there must be at least one zero eigenvalue. All singular values of $Q^{-\half} UQ^{-\half}$ are equal, thus $U = 0$. When $U$ is odd dimension, all singular values of $ Q^{-\half} UQ^{-\half}$ are equal $\| Q^{-\half} UQ^{-\half}\|_\infty$. The canonical form is block-diagonal with $2\times2$ blocks $\begin{pmatrix} 0 & u \\ -u & 0 \end{pmatrix}$ where $u=\| Q^{-\half} UQ^{-\half}\|_\infty$. 
 \\

\subsection{Form of $W$ for $T[W]=R$ in the Two-Parameter Case}
For the diagonal weight matrix $W = \begin{pmatrix} 1 & 0 \\ 0 & \omega \end{pmatrix},$
the expression for $T[W]$ becomes
\begin{equation*}
T(\omega ) = \frac{2\sqrt{\omega \det U}}{Q_{22} + \omega Q_{11}}.
\end{equation*}
Imposing the condition $T[W] = R$, we solve
$$
\frac{2\sqrt{\omega\det U}}{Q_{22} + \omega Q_{11}} = \sqrt{\frac{\det U}{\det Q}},
$$
Here $\det U \neq 0$.
This yields the quadratic equation:
\begin{equation*}
Q_{11}^2 \omega^2 - (2Q_{11}Q_{22} - 4Q_{12}^2)\omega + Q_{22}^2 = 0,
\end{equation*}
with solution
\begin{equation*}
\omega = \frac{Q_{11}Q_{22} - 2Q_{12}^2 \pm 2\sqrt{Q_{12}^2(Q_{12}^2 - Q_{11}Q_{22})}}{Q_{11}^2}.
\end{equation*}
This equation admits a unique real solution if and only if $Q_{12} = 0$, i.e. when the off-diagonal term vanishes.

For the non-diagonal weight matrix $W = \begin{pmatrix} 1 & \omega_1 \\ \omega_1 & \omega_2 \end{pmatrix},$ the expression for $T$ becomes
\begin{equation*}
T(\omega_1, \omega_2) = \frac{2\sqrt{(\omega_2 - \omega_1^2)\det U}}{Q_{22} + \omega_2 Q_{11} - 2\omega_1 Q_{12}}.
\end{equation*}
Solving the equation $T=R$ when $\det U\neq 0$:
$$
\frac{2\sqrt{(\omega_2 - \omega_1^2)\det U}}{Q_{22} + \omega_2 Q_{11} - 2\omega_1 Q_{12}} = \sqrt{\frac{\det U}{\det Q}},
$$
leads to a quadratic equation in $\omega_1$:
$$
4Q_{11}Q_{22}\omega_1^2 - 4Q_{12}(Q_{11}\omega_2 + Q_{22})\omega_1 + \left(\omega_2^2Q_{11}^2 - 2\omega_2Q_{11}Q_{22} + 4\omega_2Q_{12}^2 + Q_{22}^2\right) = 0,
$$
with $\Delta = 16(Q_{12}^2 - Q_{11}Q_{22})(Q_{11}\omega_2 - Q_{22})^2.$ This has a unique real solution if and only if $\omega_2 = \frac{Q_{22}}{Q_{11}}$, which further leads to $\omega_1 = \frac{Q_{12}}{Q_{11}}$.
\\
\subsection{A general expression of $T[W]$}
We provide a general expression for $T[W]$. Define $\{d_i\}$ and $\{\ket{d_i}\}(i=1,...,n)$  are the eigenvalues and corresponding eigenstates of $\sqrt{W} Q^{-1}$ and  $ \{ \mu_1, \mu_1, \mu_2, \mu_2, \dots, \mu_m, \mu_m, 0, \dots, 0 \} $ are the singular values of  $ U$, then $T[W]$ is expressed as
$$
T[W]=\frac{\sum^{2m}_{i,j=1}\sum^m_{k=1} 2d_i d_j \mu_k}{\sum_{i=1}^n d_i \left\langle d_i\right|\sqrt{W}\left|d_i\right\rangle}.
$$

We are interested in computing the trace norm part: $ \left\| \sqrt{W} Q^{-1} U Q^{-1} \sqrt{W} \right\|_1$.
We define $ A := \sqrt{W} Q^{-1} $, which is a real symmetric positive definite matrix. Hence, $A $ admits an eigendecomposition
$$
A = P D P^\top,
$$
where $P$ is an orthogonal matrix ($P^\top P= I$) and $D = \mathrm{diag}(d_1, \dots, d_n)$ is a diagonal matrix with strictly positive entries. The original matrix becomes
$$\sqrt{W}Q^{-1}UQ^{-1}\sqrt{W} = A U A^\top=A U A= PDP^\top UP D P^\top,$$
where $A^\top= A$ from $(P D P^\top)^\top=P D P^\top.$
Using orthogonality of $P$, we define $ \tilde{U} := P^\top UP$, so the expression becomes: $PD \tilde{U} D P^\top$.
Since trace norm (sum of singular values) is unitarily invariant, we have
$$
\left\| A U A \right\|_1 = \left\| D \tilde{U} D \right\|_1.
$$
Any real antisymmetric matrix $U \in \mathbb{R}^{n \times n}$ can be brought via an orthogonal transformation to a block-diagonal form
$$
\tilde{U} =P^\top U P = \bigoplus_{k=1}^{m} \begin{pmatrix} 0 & \mu_k \\ -\mu_k & 0 \end{pmatrix} \oplus 0_{n-2m},
$$
where $ \mu_k > 0$ and each $2 \times 2$ block corresponds to a pair of purely imaginary eigenvalues $ \pm i \mu_k$. The singular values of  $U$ are thus  $\{ \mu_1, \mu_1, \mu_2, \mu_2, \dots, \mu_m, \mu_m, 0, \dots, 0 \}$.

Suppose $D = \mathrm{diag}(d_1, \dots, d_n)$, and consider the matrix $ D \tilde{U} D$. For each $2 \times 2$ block $\begin{pmatrix} 0 & \mu_k \\ -\mu_k & 0 \end{pmatrix}$ acting on indices $i$ and $j$, the scaled version becomes
\begin{equation*}
\begin{pmatrix} d_i & 0 \\ 0 & d_j \end{pmatrix}
\begin{pmatrix} 0 & \mu_k \\ -\mu_k & 0 \end{pmatrix}
\begin{pmatrix} d_i & 0 \\ 0 & d_j \end{pmatrix}
=
\begin{pmatrix} 0 & d_i d_j \mu_k \\ -d_i d_j \mu_k & 0 \end{pmatrix}.
\end{equation*}
This new block has singular values $d_i d_j \mu_k$.
Hence, the singular values of the matrix $D \tilde{U} D$, also $\sqrt{W} Q^{-1} U Q^{-1} \sqrt{W}$, are given by
\begin{equation*}
\left\{ d_i d_j \mu_k, \ d_i d_j \mu_k, \ \dots, \ 0, \dots \right\},
\end{equation*}
with each nonzero singular value of $U$ being scaled by the product of the corresponding diagonal entries in $ D$ for the indices involved in the 2-dimensional antisymmetric block. Therefore, we express $T[W]$ as
$$T[W]=\frac{|| \sqrt{W} Q^{-1} U Q^{-1} \sqrt{W}  ||_1}{\Tr[WQ^{-1}]}=\frac{\sum^{2m}_{i,j=1}\sum^m_{k=1} 2d_i d_j \mu_k}{\sum_{i=1}^n d_i \left\langle d_i\right|\sqrt{W}\left|d_i\right\rangle}.
$$

\subsection{2-parameter Mixed qubit state estimation}
We now consider mixed probe states of the form
\begin{equation*}
\rho_0=\frac12(I+ \vec{r}_0\cdot\vec{\sigma})\,,
\end{equation*}
where $\vec{r}_0 = (r_x, r_y, r_z)$ is the Bloch vector and $\vec{\sigma} = (\sigma_1, \sigma_2, \sigma_3)$ denotes the Pauli matrices. The two parameters $\lambda_1$ and $\lambda_2$ are encoded sequentially via $U_1$ and $U_2$. After encoded and rotation, the probe state is 
\begin{equation*}
\rho_{\vec{\lambda}} = U_2 V U_1 \rho_0 U_1^\dagger V^\dagger U_2^\dagger=\frac12(I+ \vec{r}\cdot\vec{\sigma})
\end{equation*}
 where $\vec{r} = (r_x', r_y', r_z')$, with 
\begin{align*}
r_x'&=-2\kappa_2\cos\gamma  A(\varepsilon) + (1-2\kappa_2^2)B(\varepsilon)-2\kappa_1^2 \cos \varepsilon  B(0)+2\kappa_1 r_z (\kappa_2 \cos \varepsilon+ \cos \gamma \sin \varepsilon),\\
r_y'&=\cos ^2\! \gamma A(\varepsilon) \!+\! 2\kappa_1 \cos \gamma B(\varepsilon)\!+\!2\kappa_1r_z(\kappa_2 \sin \varepsilon\!-\!\cos \gamma \cos\varepsilon)\!-\!\sin^2 \!\gamma [\cos \varepsilon A(0) \!+\! \cos2\theta\sin \varepsilon B(0)],\\
r_z'&=(1-2\kappa_1^2)r_z+2\kappa_1[\cos\gamma A(0) +\kappa_2 B(0)],
\end{align*}
where
\begin{align*}
\kappa _1&=\sin\gamma \sin\theta,\quad \kappa _2=\sin\gamma \cos\theta,\\
A(\varepsilon)&=r_y \cos(\xi+\varepsilon) + r_x \sin(\xi+\varepsilon),\\
B(\varepsilon)&=r_x \cos(\xi+\varepsilon) - r_y \sin(\xi+\varepsilon).
\end{align*}
with $\xi=2\lambda_1-\phi$ and $\varepsilon=2\lambda_2+\phi.$

 
The QFIM and Uhlmann matrix are calculated by
\begin{align*}
Q_{ij}&= (\partial _i\vec{r})\cdot(\partial _j\vec{r})+\frac{(\vec{r}\cdot\partial _i\vec{r})(\vec{r}\cdot\partial _j\vec{r})}{1-|\vec{r}|^2},\\
U_{ij}&=\vec{r}\cdot (\partial _i\vec{r}\times \partial _j\vec{r}).
\end{align*}

Setting $\lambda_2 = 0$ simplifies $Q_{22}$ without loss of generality, 
the QFIM components for parameters $\lambda_1, \lambda_2$ are
\begin{align*}
Q_{11} &= 4(r_x^2 + r_y^2),\\
Q_{12} &= Q_{21}=4 \Big( (r_x^2 + r_y^2) (1-2\kappa_1^2)- 2r_z \kappa_1\left[\cos\gamma A(0) +\kappa_2 B(0)\right] \Big), \\
Q_{22} &=(\partial _2r_x)^2+(\partial _2r_y)^2,
\end{align*}
where
\begin{align*}
\partial _2r_x&= 2\kappa_1r_z(\cos \gamma \cos\phi-\kappa_2 \sin \phi\!)-\cos2 \gamma\cos\phi A(0)+[(2\kappa_2^2-1)\sin \phi-2\kappa_1\cos \gamma]B(0) ,\\
\partial _2r_y&=\cos^2\! \gamma B(\phi)\!-\!2 \kappa_2 \cos\! \gamma \!+\! 2\kappa_1r_z(\cos \gamma \sin\phi \!+\! \kappa_2 \cos \phi) \! +\! \sin^2 \!\gamma \sin\!\phi A(0)\!+\!(\kappa_1^2\!-\!\kappa_2^2)\cos\!\phi B(0).
\end{align*}

The Uhlmann matrix components are
\begin{align*}
U_{11} &= U_{22} = 0, \\
U_{12} &= -U_{21} = 8 |\vec{r}|^2 \kappa_1 \left[\kappa_2 A(0)-\cos\gamma B(0)\right].
\end{align*}

\subsection*{Two-parameter estimation for SU(2) unitary encoding}

We begin by considering the two-parameter estimation problem, where the generator takes the form
\begin{equation*}
    H_{B, \theta} = H = B\left( \cos\theta J_x + \sin\theta J_z \right) = B J_{\vec{n}_\theta}
\end{equation*}
with $J_x, J_y, J_z$ being generators of the $SU(2)$ Lie algebra satisfying the commutation relations:
\begin{align*}
    [J_x, J_y] &= i J_z, \\
    [J_y, J_z] &= i J_x, \\
    [J_z, J_x] &= i J_y.
\end{align*}

We adopt the shorthand notation $J_{\vec{n}_\theta} = \vec{n}_\theta \cdot \vec{J}$, where $\vec{n}_\theta = (\cos\theta, 0, \sin\theta)$ is a unit vector and $\vec{J} = (J_x, J_y, J_z)$ denotes the generator vector.

Next, we compute the QFIM and the Uhlmann matrix. Note that we do not fix the Hilbert space dimension; we only assume that the generators form an $SU(2)$ algebra.

The QFIM entries are given by
\begin{align*}
    Q_{BB} &= 4 t^2 \left( \langle J_{\vec{n}_\theta}^2 \rangle_0 - \langle J_{\vec{n}_\theta} \rangle_0^2 \right),\\
    Q_{\theta\theta} &= 16 \sin^2 \frac{Bt}{2} \left( \langle J_{\vec{n}_1}^2 \rangle_0 - \langle J_{\vec{n}_1} \rangle_0^2 \right), \\
    Q_{B\theta} &= -4 t \sin \frac{Bt}{2}  \left( \left\langle \{ J_{\vec{n}_1}, J_{\vec{n}_\theta} \} \right\rangle_0 - 2 \langle J_{\vec{n}_1} \rangle_0 \langle J_{\vec{n}_\theta} \rangle_0 \right).
\end{align*}

The nontrivial Uhlmann matrix element reads
\begin{equation}
    U_{B\theta} = 4t \sin\frac{Bt}{2} \langle J_{\vec{n}_2} \rangle_0,
\end{equation}
with 
\begin{align*}
    &\vec{n}_\theta = (\cos \theta, 0, \sin \theta),\\
    &\vec{n}_1 = (\cos\frac{Bt}{2}\sin\theta,-\sin\frac{Bt}{2},-\cos\frac{Bt}{2}\cos\theta),\\
    &\vec{n}_2 = \vec{n}_\theta \times \vec{n}_1 = (\sin\frac{Bt}{2}\sin\theta,\cos\frac{Bt}{2},-\sin\frac{Bt}{2}\cos\theta).
\end{align*}

In the case of qubit probes, the QFIM entries are
\begin{align*} \label{2qubitQ}
  &Q_{BB} = t^2 [1-(\vec{n}_\theta \cdot \vec{r}_0)^2],\\ 
  &Q_{\theta\theta} = 4 \sin^2 \frac{Bt}{2} [1-(\vec{n}_1\cdot\vec{r}_0)^2],\\ 
    &Q_{B\theta} = 2t\sin\frac{Bt}{2}(\vec{n}_1 \cdot \vec{r}_0)(\vec{n}_\theta \cdot \vec{r}_0),
\end{align*}
and the Uhlmann matrix off-diagonal entry is
\begin{equation*}\label{2qubitU}
  U_{\theta B} = 2t\sin\frac{Bt}{2}\vec{n}_2\cdot\vec{r}_0.
\end{equation*}
\\

\subsection{Three-parameter qutrit estimation for SU(2) unitary encoding}
Having studied the two-parameter case in the previous section, we now consider the compatibility of estimating three parameters in an $SU(2)$ unitary encoding. The generator takes the form
\begin{equation*}
    H_{B, \theta, \varphi} = H = B J_{\vec{n}^{(3)}_\theta}
\end{equation*}
where $J_x, J_y, J_z$ are elements of the $SU(2)$ algebra. We use the notation
\[
J_{\vec{n}^{(3)}_\theta} = \vec{n}_\theta \cdot \vec{J}
\]
with the direction vector
\[
\vec{n}_\theta^{(3)} = (\cos\theta \cos\varphi, \cos\theta \sin\varphi, \sin\theta)
\]
and $\vec{J} = (J_x, J_y, J_z)$.

Following the same reasoning as in the two-parameter case, we now compute the QFIM and Uhlmann matrix. For the parameter $\theta$, we obtain
\begin{equation*}
    \partial_\theta H = B J_{\vec{n}_{\theta ^\prime}^{(3)}}
\end{equation*}
with
\[
\vec{n}_{\theta ^\prime}^{(3)} = (-\sin\theta \cos\varphi, -\sin\theta \sin\varphi, \cos\theta).
\]
Define $\vec{n}_0 = (-\sin\varphi, \cos\varphi, 0)$, and after some algebra we find
\begin{equation*}
    H^\times(J_{\vec{n}_0}) = [H, J_{\vec{n}_0}] = iB J_{\vec{n}_{\theta ^\prime}^{(3)}} = i \partial_\theta H.
\end{equation*}
Proceeding as before, we obtain
\begin{equation*}
    \mathcal{H}_\theta = 2 \sin\frac{Bt}{2} J_{\vec{n}_1^{(3)}}
\end{equation*}
where the unit vector $\vec{n}_1^{(3)}$ is defined as
\begin{equation*}
    \vec{n}_1^{(3)} =
    \begin{pmatrix}
  \sin  \frac{Bt}{2} \sin\varphi + \cos\frac{Bt}{2} \sin\theta \cos\varphi \\
    -\sin \frac{Bt}{2} \cos\varphi + \cos\frac{Bt}{2} \sin\theta \sin\varphi \\
    -\cos\frac{Bt}{2} \cos\theta
    \end{pmatrix}.
\end{equation*}

In this case
\begin{equation*}
    \mathcal{H}_\varphi =  2 \cos\theta \sin\frac{Bt}{2} J_{\vec{n}_{2}^{(3)}}, \,\,\,\,    \mathcal{H}_B= -t J_{\vec{n}_{\theta}^{(3)}}
\end{equation*}
with \begin{equation*}
   \vec{n}_{2}^{(3)}=  \begin{pmatrix}
  \cos  \frac{Bt}{2} \sin\varphi - \sin\frac{Bt}{2} \cos\varphi \\
    -\cos \frac{Bt}{2} \cos\varphi - \sin\frac{Bt}{2} \sin\theta \sin\varphi \\
    \sin\frac{Bt}{2} \cos\theta
    \end{pmatrix}.
\end{equation*}

The QFIM entries are given by
\begin{align*}
Q_{BB} &= 4 t^2 \left( \langle J_{\vec{n}_\theta^{(3)}}^2 \rangle_0 - \langle J_{\vec{n}_\theta^{(3)}} \rangle_0^2 \right), \\
Q_{\theta\theta} &= 16 \sin^2\frac{Bt}{2} \left( \langle J_{\vec{n}_1^{(3)}}^2 \rangle_0 - \langle J_{\vec{n}_1^{(3)}} \rangle_0^2 \right), \\
Q_{\varphi\varphi} &= 16 \cos^2\theta \sin^2\frac{Bt}{2} \left( \langle J_{\vec{n}_2^{(3)}}^2 \rangle_0 - \langle J_{\vec{n}_2^{(3)}} \rangle_0^2 \right),\\
Q_{B\theta} &= -4t \sin\frac{Bt}{2} \left( \langle \{ J_{\vec{n}_\theta^{(3)}}, J_{\vec{n}_1^{(3)}} \} \rangle_0 - 2 \langle J_{\vec{n}_\theta^{(3)}} \rangle_0 \langle J_{\vec{n}_1^{(3)}} \rangle_0 \right), \\
Q_{B\varphi} &= -4t \cos\theta \sin\frac{Bt}{2} \left( \langle \{ J_{\vec{n}_\theta^{(3)}}, J_{\vec{n}_2^{(3)}} \} \rangle_0 - 2 \langle J_{\vec{n}_\theta^{(3)}} \rangle_0 \langle J_{\vec{n}_2^{(3)}} \rangle_0 \right), \\
Q_{\theta\varphi} &= 8 \cos\theta \sin^2\frac{Bt}{2} \left( \langle \{ J_{\vec{n}_1^{(3)}}, J_{\vec{n}_2^{(3)}} \} \rangle_0 - 2 \langle J_{\vec{n}_1^{(3)}} \rangle_0 \langle J_{\vec{n}_2^{(3)}} \rangle_0 \right).
\end{align*}

The non-trivial Uhlmann matrix elements are
\begin{align*}
U_{B\theta} &= -4t \sin\frac{Bt}{2} \langle J_{\vec{n}_\theta^{(3)} \times \vec{n}_1^{(3)}} \rangle_0, \\
U_{B\varphi} &= -4t \cos\theta \sin\frac{Bt}{2} \langle J_{\vec{n}_\theta^{(3)} \times \vec{n}_2^{(3)}} \rangle_0, \\
U_{\theta\varphi} &= 8 \cos\theta \sin^2\frac{Bt}{2} \langle J_{\vec{n}_1^{(3)} \times \vec{n}_2^{(3)}} \rangle_0.
\end{align*}

We choose the initial state  
\begin{equation*}
\ket{\psi_0} =\cos\frac{\alpha}{2} \ket{1} + e^{i\beta} \sin\frac{\alpha}{2} \ket{-1}.
\end{equation*}
We get the entries of QFIM and Uhlmann matrix are 
\begin{align*}
Q_{BB}&=2t^2\left(1 +\sin^2\theta \cos\!2\alpha + \cos^2\theta X  \right), \\
Q_{\theta\theta}& = 8 \sin^2\! \frac{Bt}{2} \!\left( 1 \!- \!\cos^2\!\frac{Bt}{2} \cos\!2\alpha \cos^2\!\theta + \left[\cos^2\!\frac{Bt}{2} \sin^2\!\theta-\sin^2\!\frac{Bt}{2}\right]X-\sin\!Bt \sin\!\theta Y\big]\! \right),\\
Q_{\varphi\varphi} &=8 \cos^2\!\theta  \sin^2\!\frac{Bt}{2} \left( 
1\! - \! \sin^2\!\frac{Bt}{2}\cos\!2\alpha\cos^2\!\theta-\left[\cos^2 \!\frac{Bt}{2}-\sin^2\!\theta\sin^2 \!\frac{Bt}{2}\right]X+ \sin\!Bt \sin\!\theta Y \right),\\
Q_{B\theta}&=Q_{\theta B} =4t \cos\theta\sin\!\frac{Bt}{2}\Big( \sin\!\frac{Bt}{2}Y- \cos\!\frac{Bt}{2}\sin\theta \left[ \cos2\alpha + X \right]  \Big),\\
Q_{B\varphi}&= Q_{\varphi B}=4t \cos^2\theta \sin\!\frac{Bt}{2}\Big( \sin\!\frac{Bt}{2} \sin\!\theta \big[\cos\!2\alpha+X \big]  + \cos\!\frac{Bt}{2} Y \Big),\\
Q_{\theta\varphi} &= 4\cos\theta\sin^2 \!\frac{Bt}{2} \Big( \sin Bt \left[ \cos\!2\alpha \cos^2\!\theta-(\sin^2\!\theta+1)X \right]- 2 \cos Bt \sin\!\theta Y\Big),
\end{align*}
with $X=\sin\!\alpha \cos(\beta\!-\!2\varphi),$ and $Y=\sin\!\alpha \sin(\beta\!-\!2\varphi).$ 

The entries of Uhlmann matrix are 
\begin{align*}
U_{B\theta} &= -U_{\theta B} =4t \cos\alpha \cos\theta \sin^2 \!\frac{Bt}{2},\\
U_{B\varphi} &= -U_{\varphi B} =2t \cos\alpha \cos^2\!\theta \sin Bt,\\
U_{\theta\varphi} &=-U_{\varphi\theta} = -4 \cos\alpha \sin^2\! \frac{Bt}{2} \sin2\theta.
\end{align*}

Because all $\beta$ appears in form $\beta\!-\!2\varphi$, setting $\beta=0$ would not change the optimal precision. Then the SLD operators are
\begin{align*}
[L_B]_{11}& =[L_B]_{22}=[L_B]_{33}=0,\\
[L_B]_{12}&=[L_B]_{21}^*= \sqrt{2}\cos\frac{\alpha}{2}\cos\theta\Big(i e^{i\varphi}\sin\frac{\alpha}{2}+\cos\frac{\alpha}{2}(i\cos\varphi+\sin\varphi)\Big) ,\\
[L_B]_{13}&=[L_B]_{31}^*=  -2i\sin\alpha\sin\theta,\\
[L_B]_{23}&=[L_B]_{32}^*= -i \sqrt{2} e^{-i\varphi}\cos\theta\sin\frac{\alpha}{2}\Big( e^{2i\varphi}\cos\frac{\alpha}{2} +\sin\frac{\alpha}{2}\Big),\\
[L_\theta]_{11}& =[L_\theta]_{22}=[L_\theta]_{33}=0,\\
[L_\theta]_{12}&=[L_\theta]_{21}^*=\sqrt{2} \cos\frac{\alpha}{2} \Big(-\!e^{-i\varphi}\cos\frac{\alpha}{2}[\sin\!\frac{Bt}{2}-i\cos\!\frac{Bt}{2}\sin\theta]+e^{i\varphi}\sin\frac{\alpha}{2}[\sin\!\frac{Bt}{2}+i\cos\!\frac{Bt}{2}\sin\theta]\Big),\\
[L_\theta]_{13}&=[L_\theta]_{31}^*= 2i\cos\!\frac{Bt}{2}\cos\theta\sin\alpha,\\
[L_\theta]_{23}&=[L_\theta]_{32}^*= \sqrt{2}\sin\frac{\alpha}{2}\Big(e^{-i\varphi}\sin\frac{\alpha}{2}[\sin\!\frac{Bt}{2}-i\cos\!\frac{Bt}{2}\sin\theta]-e^{i\varphi}\cos\frac{\alpha}{2}[\sin\!\frac{Bt}{2}+i\cos\!\frac{Bt}{2}\sin\theta]\Big),\\
[L_\varphi]_{11}& =[L_\varphi]_{22}=[L_\varphi]_{33}=0,\\
[L_\varphi]_{12}&=[L_\varphi]_{21}^*=\sqrt{2} e^{-i\varphi}\cos\frac{\alpha}{2}\Big(
e^{2i\varphi}\sin\frac{\alpha}{2}\big[\cos\frac{Bt}{2}-i\sin\frac{Bt}{2}\sin\theta\big]
-\cos\frac{\alpha}{2}\big[\cos\frac{Bt}{2}+i\sin\frac{Bt}{2}\sin\theta\big]
\Big),\\
[L_\varphi]_{13}&=[L_\varphi]_{31}^*= -2i\sin\alpha\cos\theta\sin\!\frac{Bt}{2},\\
[L_\varphi]_{23}&=[L_\varphi]_{32}^*=\sqrt{2} e^{-\!i\varphi}\sin\frac{\alpha}{2}\Big(
- e^{2i\varphi}\cos\frac{\alpha}{2}\big[\cos\frac{Bt}{2}-i\sin\frac{Bt}{2}\sin\theta\big]
+ \sin\frac{\alpha}{2}\big[\cos\frac{Bt}{2}+i\sin\frac{Bt}{2}\sin\theta\big]
\Big).
\end{align*}

 Then we apply Theorom~\ref{CH2} to calculate $C_H$. When $\alpha=\pi/4$, $t=1,$  $B=\pi,$  $\theta=0,$  $\varphi=0$,
 based on the inner product $\langle L_{\nu},P_j\rangle_{\rho_{\vec{\lambda}}}=0$ and $\Tr[\rho_{\vec{\lambda}}P_j]=0$, we obtain
$$
P_1 = 
\begin{pmatrix}
\frac{1}{\sqrt{3}} & 0 & 0 \\
0 & \frac{1}{\sqrt{3}} & 0 \\
0 & 0 & -\frac{2}{\sqrt{3}}
\end{pmatrix}, \quad
P_2 =
\begin{pmatrix}
0 & -\dfrac{i\sqrt{2}}{2+\sqrt{2}} & 0 \\
\dfrac{i\sqrt{2}}{2+\sqrt{2}} & 0 & -i \\
0 & i & 0
\end{pmatrix},
$$
$$
P_3 =
\begin{pmatrix}
0 & 1-\sqrt{2} & 0 \\
1-\sqrt{2} & 0 & 1 \\
0 & 1 & 0
\end{pmatrix}, \quad
P_4 =
\begin{pmatrix}
0 & 0 & 1 \\
0 & 0 & 0 \\
1 & 0 & 0
\end{pmatrix}, \quad
P_5 =
\begin{pmatrix}
1 & 0 & 0 \\
0 & -1 & 0 \\
0 & 0 & 0
\end{pmatrix}.
$$

The adjusting operator $P_{\vec{\lambda}}$ with $[P_{\vec{\lambda}}]_{i,j}=\Tr[\rho_{\vec{\lambda}}P_iP_j],$ is given by
$$
P_{\vec{\lambda}} =
\begin{pmatrix}
\dfrac{1}{6}\left(5 - \tfrac{3}{\sqrt{2}}\right) & 0 & 0 & -\tfrac{1}{2\sqrt{6}} & \dfrac{\cos^2(\pi/8)}{\sqrt{3}} \\
0 & 0 & 0 & 0 & 0 \\
0 & 0 & 0 & 0 & 0 \\
-\tfrac{1}{2\sqrt{6}} & 0 & 0 & 1 & \tfrac{1}{2\sqrt{2}} \\
\dfrac{\cos^2(\pi/8)}{\sqrt{3}} & 0 & 0 & \tfrac{1}{2\sqrt{2}} & \cos^2(\pi/8)
\end{pmatrix},
$$
and the matrix $L_{\vec{\lambda}}$ is
$$
L_{\vec{\lambda}} =
\begin{pmatrix}
0 & -\sqrt{2} & 0 \\
\sqrt{2} & 0 & 0 \\
0 & 0 & 0
\end{pmatrix}.
$$

The matrix $S$ is defined by 
$[S]_{ij} = \text{Im} \Tr(\rho_{\vec{\lambda}} L_i P_j),$ and in our case takes the form
$$
S = 
\begin{pmatrix}
0 & 0 & 0 & 0 & 0 \\
0 & 0 & 0 & 0 & 0 \\
\dfrac{\sqrt{3}}{2} & 0 & 0 & -1 & \tfrac{1}{2}
\end{pmatrix}.
$$

Finally minimize bound with $5\times3$ matrix $K$ and obtain $C_H=C_T=\frac{11+\sqrt2}{8}\approx1.5518$ with $K=\mathbf{0}$. 

\bibliographystyle{plain}
\bibliography{wbb}

\end{document}